\documentclass[prb, twocolumn, showpacs, amsmath, amssymb,superscriptaddress]{revtex4-1}
\usepackage{amssymb,amsfonts,amsmath} 
\usepackage{graphicx}
\usepackage{xcolor}
\usepackage{hyperref}
\usepackage{longtable}
\usepackage{bbm}
\usepackage{ulem}
\usepackage{enumitem}
\usepackage{bm}
\usepackage{braket}
\usepackage{comment}
\usepackage{cleveref}
\usepackage{cancel}
\newcommand{\sgn}{\mathrm{sgn}}

\renewcommand{\Im}{\operatorname{Im}}

\usepackage{nicefrac}

\begin{document} 

\title{Dynamical instability in a Floquet-Driven Dissipative System}

\author{Takuya Okugawa} 
\email{to2462@columbia.edu}
\affiliation{Department of Physics, Columbia University, New York, NY 10027, USA}
\affiliation{Max Planck Institute for the Structure and Dynamics of Matter, Center for Free Electron Laser Science, 22761 Hamburg, Germany.}

\author{Jens Paaske}
\affiliation{Center for Quantum Devices, Niels Bohr Institute, University of Copenhagen, 2100 Copenhagen, Denmark}

\author{Martin Eckstein}
\affiliation{Institute of Theoretical Physics, University of Hamburg, 20355 Hamburg, Germany}
\affiliation{The Hamburg Centre for Ultrafast Imaging, Hamburg, Germany}

\author{Michael A. Sentef}
\affiliation{Institute for Theoretical Physics and Bremen Center for Computational Materials Science, University of Bremen, 28359 Bremen, Germany}
\affiliation{Max Planck Institute for the Structure and Dynamics of Matter, Center for Free Electron Laser Science, 22761 Hamburg, Germany.}

\author{Angel Rubio}
\affiliation{Max Planck Institute for the Structure and Dynamics of Matter, Center for Free Electron Laser Science, 22761 Hamburg, Germany.}
\affiliation{Center for Computational Quantum Physics, Simons Foundation Flatiron Institute, New York, NY 10010 USA.}
\affiliation{Nano-Bio Spectroscopy Group, Departamento de Fisica de Materiales, Universidad del Pa\'is Vasco, UPV/EHU- 20018 San Sebasti\'an, Spain.}

\author{Andrew J. Millis}
\affiliation{Department of Physics, Columbia University, New York, NY 10027, USA}
\affiliation{Center for Computational Quantum Physics, Simons Foundation Flatiron Institute, New York, NY 10010 USA.}

\begin{abstract} 
We analyse the magnon spectrum and distribution function of the antiferromagnetic phase of the Floquet-driven Hubbard model. Above a critical drive strength, we find a dynamical instability, resulting from a change in sign of the magnon damping at a non-zero wavevector. The change in sign means that infinitesimal fluctuations grow with time, corresponding to an instability of the driven state.  Implications for the nonequilibrium distribution function and the strong drive nonlinear dynamics are discussed. 
\end{abstract}

\date{\today} 
\maketitle

\section{Introduction}
Recent progress in laser technology is enabling the control of material properties on demand~\cite{Basov17,Torre2021,Murakami2025} by dynamically changing phases and inducing states of matter that are inaccessible under equilibrium conditions~\cite{Oka_2019}. Experimentally reported phenomena include insulator-metal transitions \cite{Pashkin11,Forst15,Scarlatella2019,Stoica22,Xu22,Verma24}, 
optically induced chirality and gap formation~\cite{McIver_2019}, light-induced high transition temperature superconductivity~\cite{Cavalleri2017}, polaritonic bose condensates, laser-controlled magnetism~\cite{Kirilyuk_2010}, and polarization switching in ferroelectrics \cite{Li19}.

As discussed in the context of light-controlled magnets\cite{Kirilyuk_2010}, materials control can arise from distribution function changes, where a population of excitations is created, either by direct excitation~\cite{Szymaifmmode2006,Szymaifmmode2007,Wouters2007,Pashkin11,Forst15,Scarlatella2019,Stoica22,Xu22,Verma24} or parametric amplification \cite{Babadi17,Murukami17} and the resulting non-linear interactions change the state~\cite{Szymaifmmode2006,Szymaifmmode2007,Wouters2007,Pashkin11,Forst15,Scarlatella2019,Stoica22,Xu22,Verma24}, or from drive-induced changes to the Hamiltonian \cite{Cavalleri2017,McIver_2019}. In this paper we uncover a different route in a Floquet-driven magnon system: a nonequilibrium-driven change in sign of mode damping, so that modes that were previously stable become unstable, growing in time and driving the system to a new state presumably characterized by a spatial and temporal pattern similar to that occurring in classical driven dissipative systems~\cite{Cross1993}. This may be viewed as a gain instability somewhat analogous to that occurring in a laser, but here created by a Floquet drive rather than an explicit electronic population inversion. A gain instability towards a stable masing state induced by nonequilibrium drive has been observed in microwave resonators with voltage-biased quantum dots~\cite{Liu2015Jan, Stehlik2016Nov} serving as the gain medium, and the physics of a nonequilibrium-induced change of sign of dissipation has recently been discussed in the context of nonequilibrium field theories of driven dissipative systems \cite{Daviet2024,Zelle25,Hermansen2025Nov}.

Here, we study the effects of a 'Floquet' (time periodic) drive on a lattice model of interacting electrons with an insulating antiferromagnetic ground state, minimally coupled  to monochromatic radiation at a frequency much higher than any system excitation energy. A weak tunnel coupling to a broad-band electronic reservoir prevents runaway heating and, as will be seen, enables an interesting gain mechanism. The low energy excitations of the system are antiferromagnetic magnons, weakly damped by the coupling to the electronic reservoir and we find that sufficiently large drive produces a dynamical instability at which the magnon damping at a certain non-zero wavevector changes sign. The instability originates from Floquet drive-enabled processes that are forbidden in equilibrium and, above a critical drive strength, lead to a gain: an exponential growth in time of magnon modes in response to perturbations of particular wavevectors. This exponential growth corresponds to a spontaneous emission of magnons stimulated by application of an external field, and is in some sense analogous to a laser, although at a non-zero wavevector and in a magnetic system.  These findings correct previous work on this model by some of the authors that reported a $q=0$ instability interpreted as a divergence of a distribution function at zero wavevector \cite{Walldorf2019} and provide a microscopic confirmation of the conjecture of Zelle et al~\cite{Zelle2024} that 
instabilities of driven magnon systems may be understood in terms of change in sign of the dissipation.

This paper is organized as follows. In \Cref{sec:model} we present the model and define key quantities used in this work. In \Cref{sec:Result_discussion} we present the main results of our numerical simulations and discuss the physical consequences. In \Cref{sec:_of_the_inetability} we provide analytical arguments demonstrating the physical origin of the findings. In \Cref{sec:Magnon_distribution_function} we discuss the implications of our findings for the magnon distribution function. Finally, \Cref{sec:conclusion} provides a summary and conclusion. Supplementary material, including detailed analytic derivations and discussion of numerics, is presented in Reference \onlinecite{SM}.

\section{Model and method}
\label{sec:model}

We study the Floquet-driven Hubbard model of electrons coupled by a site-local interaction and connected to a particle reservoir and driven by an electric field \cite{Walldorf2019} described by a spatially uniform vector potential $\vec{A}(t)$: 
\begin{equation}
H=\sum_{\bm{k}\sigma}\epsilon_{\bm{k}}(t)\hat{c}^\dagger_{\bm{k}\sigma}\hat{c}_{\bm{k}\sigma}+U\sum_i\hat{n}_{i\uparrow}\hat{n}_{i\downarrow}+H_{\text{res}}
\label{eq:H}
\end{equation}
with $\hat{n}_{i\sigma}$ the density of particles of spin $\sigma$ on site $i$ and $\epsilon_{\bm{k}}(t)=-2\Tilde{t} \sum_{l=x,y}\mathrm{cos}\left(k_l a_l+ A_l(t) \right)$, where $\Tilde{t}$ is the hopping parameter\footnote{To distinguish time variable $t$, we use $\Tilde{t}$ for the hopping parameter.} and we set $\hbar=e=a=k_B=1$ with $a$ being the lattice constant. Throughout the paper, we express all energies in units of $\Tilde{t}$.

We consider a homogeneous monochromatic drive at frequency $\Omega$ and will be interested in $\Omega$ larger than the relevant scales of the system part of $H$. In the main text, we consider linearly polarized radiation with $A_l(t) =-\frac{E \mathrm{sin}(\Omega t)}{\Omega} $ unless otherwise specified. $H_{\text{res}}=g\sum_{\bm{k}\sigma}\hat{c}^\dagger_{\bm{k}\sigma}a_{\bm{k}\sigma}+\text{H.c.}$ gives the coupling to a reservoir (electronic states created by $a^\dagger_{\bm{k}\sigma}$) which we take to be noninteracting and to have a constant density of states $N_{\text{res}}$ with a bandwidth much larger than any other scale in the problem. The reservoir is held at a constant temperature which in our numerical calculations is set to $T=0.01$. The reservoir coupling gives rise to an electron lifetime (contribution to imaginary part of electron self energy) $\Gamma=g^2N_{\text{res}}$ that is frequency-independent at the scales of interest here. The reservoir system can either be thought of as an actual electronic substrate or an additional electronic band, or as a means of emulating a minimal model for phonon induced damping.

At half filling (one electron per site), the equilibrium ground state has a two-sublattice (wavevector $\bm{Q}=(\pi,\pi)$) magnetic order that fully gaps the electronic spectrum. The low energy excitations are magnons and we study the effect of the drive on the magnon spectrum and distribution. Following previous work \onlinecite{Mitra2006,Walldorf2019}, we use a Keldysh field theory formalism in a mean field plus one loop approximation to compute the retarded ($R$), advanced ($A$) and Keldysh ($K$) components of the magnetic response function $\chi$. The mean field plus one-loop approximation is known \cite{Schrieffer89} to reproduce with semiquantitative accuracy the equilibrium zero temperature magnon spectrum of this model. The spatial symmetry breaking implied by the two-sublattice magnetic order means that in the magnetic Brillouin zone the response function $\bm{\chi}$ is a $2\times 2$ matrix with diagonal components representing the uniform and staggered components of the magnetization and the off diagonal components giving the coupling between them. Each component is also a matrix in the Floquet indices. 

Within the random phase approximation (RPA), the components of the response function describing fluctuations transverse ($\bot$) to the direction of the ground state order parameter are
\begin{align}
\bm{\chi}^{\bot, R/A}(\omega)& \equiv \left[\frac{1}{2I}-\bm{\Pi}^{\bot, R/A} (\omega) \right]^{-1} \label{eq:chiRA}\\ 
\bm{\chi}^{\bot, K}(\omega) & \equiv \Biggl( \bm{\chi}^{\bot, R}(\omega) \bm{\Pi}^{\bot, K} (\omega) \bm{\chi}^{\bot, A}(\omega)  \Biggr) 
\end{align}
where $I=U/3$. 

The $\bm{\Pi}$ are electron polarization bubbles defined in the RPA as convolutions of the electronic propagators computed in the mean field ground state in the presence of the drive field and dissipation. For later convenience, we make the sublattice matrix and Floquet index structure explicit as
\begin{equation}
\bm{\Pi}^{\bot, R/A/K}=
\begin{pmatrix}
\Pi^{\bot, R/A/K}_{0,\bm{q},mn} & \Pi^{\bot, R/A/K}_{Q,\bm{q},mn} \\
\Pi^{\bot, R/A/K}_{Q,\bm{q},mn} & \Pi^{\bot, R/A/K}_{0, \bm{Q}+\bm{q},mn} \\
\end{pmatrix} \label{eq:R_pi_matrix}
\end{equation} 
with the first entry of the subscript taking the values $0$ for the diagonal uniform-uniform and staggered-staggered components of the polarizibility and $Q$ for the cross coupling terms involving scattering by the antiferromagnetic wavevector $Q$, the second entry denoting the momentum $\bm{q}$ in the reduced Brillouin zone with $\bm{Q}+\bm{q}$ denoting fluctuations of the staggered magnetization, and the third and fourth entries $mn$ denoting the Floquet indices ($|n|<n_{\text{max}}$, where $n_{\text{max}}$ is a Floquet cutoff). We are interested in fluctuations of the staggered magentization with small momenta $\bm{q}$, i.e. fluctuations with wave vectors  close to $\bm{Q}=(\pi,\pi)$ in the original Brillouin zone.

We perform high resolution numerical computations of the components of $\bm{\Pi}$. In the RPA used here, the frequency integrals may be performed analytically. For the momentum summation, we use a $100 \times 100$ ${\bf k}-$mesh over the original Brillouin zone and select ${\bf k}-$points within the magnetic Brillouin zone. Importantly for our results, we use a fine-grid sampling of magnon frequency $\omega_q$ in the low $\omega_q$ regime. We complement the numerical studies with a theoretical analysis valid for  $\Omega > U >\tilde{t}$ and in the $\Gamma \ll \tilde{t}$-limit. The technical details are given in the supplementary material in Reference \onlinecite{SM}.

\section{Result and discussion}
\label{sec:Result_discussion}
\begin{figure}[t!]
\begin{center} 
\includegraphics[width=0.5\textwidth, angle=-0]{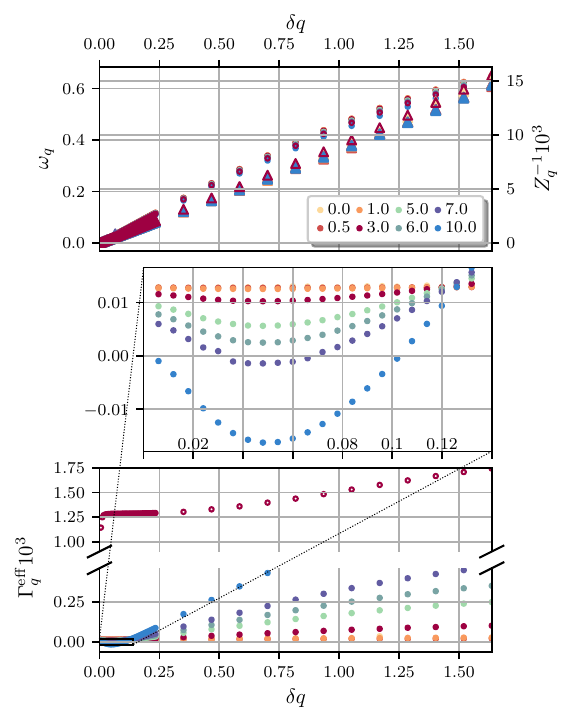}
\caption{
The upper panel shows magnon frequency $\omega_q$ (circle) as well as the inverse spectral weight $Z_q^{-1}$ (triangle) as a function of $\delta q=|\bm{Q}-\bm{q}|$ with $\bm{Q}=\left(\pi,\pi\right)$ for several drive amplitudes $E$, which are indicated in the legend on top of the upper panel. The lower panel displays the effective damping $\Gamma_{q}^{\text{eff}}$, and the middle panel provides a zoom-in of the lower panel for small $\delta q$ regime. The parameters are $I=5$, $T=0.01$, $\Omega=30$, $\Gamma=0.02$, and $n_{\text{max}}=1$. The $\Gamma=0.2$ case is also plotted as empty circles.
}
\label{fig:Lorenzian_fit}
\end{center}
\end{figure}

\subsection{Imaginary Part of the Fluctuation Propagator}
\label{subsec:imaginary_chi}

The important physical excitations are obtained from the $0-0$ Floquet components of the retarded response function $\bm{\chi}^R$. We find in agreement with previous works~\cite{Schrieffer89,Walldorf2019} that at frequencies small compared to the electronic gap, the component of $\bm{\chi}^R$ pertaining to the transverse fluctuations of the staggered magnetization at wavevectors near the ordering wavevector is well approximated by a very weakly damped pole
\begin{equation}
\chi^{\bot, R}_{0,\bm{q},00}(\omega) \approx \frac{Z_q}{\omega_q-\omega-i\Gamma_q^{\text{eff}}}.
\label{eq:fit_chi}
\end{equation}
We fit our numerically calculated $\text{Im}\left[\chi^{\bot, R}_{0,\bm{q},00}(\omega)\right]$ to \Cref{eq:fit_chi}, extracting $Z_q$ and $\Gamma_{q}^{\text{eff}}$ ($\omega_q$ is identified from the pole of $\text{Im}\left[\chi^{\bot, R}_{0,\bm{q},00}(\omega)\right]$). The $Z_q$ agrees with a direct computation (i.e. $Z_q =\int \frac{d\omega}{\pi}~ \text{Im}\left[\chi^{\bot, R}_{0,\bm{q},00}(\omega)\right]$). Representative results are given in \Cref{fig:Lorenzian_fit}. The upper panel shows the magnon frequency $\omega_q$ alongside the inverse spectral weight $Z_q^{-1}$ for several drive amplitudes. 
Both $\omega_q$ and $Z_q^{-1}$ vary linearly with $\delta q$ and are nearly independent of the drive amplitude $E$ and reservoir coupling $\Gamma$, in agreement with previous results~\cite{Schrieffer89,Walldorf2019}. The modest changes from the $E=0$ (non-driving) values can be understood via the high drive frequency renormalization of the hopping parameter $\Tilde{t} \rightarrow \Tilde{t} J_0(E/\Omega)$. However, the spin wave damping $\Gamma_{q}^{\text{eff}}$ exhibits a strong dependence on $E$ and $\Gamma$, as shown in the lower panel. At low $E$, $\Gamma_{q}^{\text{eff}}$ is essentially $q$-independent. As the drive amplitude $E$ is increased, $\Gamma_{q}^{\text{eff}}$ acquires a momentum-dependent structure whose details depend on the relative strength of the reservoir coupling and the drive. The middle panel displays an expanded view $\Gamma^{\text{eff}}$; we see that as the drive strength increases, a minimum appears at a nonzero $\delta q^\star$. As the drive strength is further increased, $\Gamma_{q}^{\text{eff}}$ changes sign, becoming negative for a range of $\delta q$  around $\delta q^\star$. This sign change implies that small fluctuations of the magnetization in this wavevector regime grow (see Reference \onlinecite{SM} for the detail), signaling a dynamical instability of the magnetic state~\cite{Szymaifmmode2006, Szymaifmmode2007, Hanai2017, Scarlatella2019, Wouters2007}. Because the instability appears at one loop order in the retarded response function, it arises from a drive-induced instability in the actual dynamics of the system, in contrast to other cases~\cite{Pashkin11,Forst15,Stoica22,Xu22,Verma24,Babadi17,Murukami17} where instabilities arise from the combination of a drive induced excess in the magnon population and magnon-magnon interactions.

\begin{figure}[t!]
\begin{center} 
\includegraphics[width=0.45\textwidth, angle=-0]{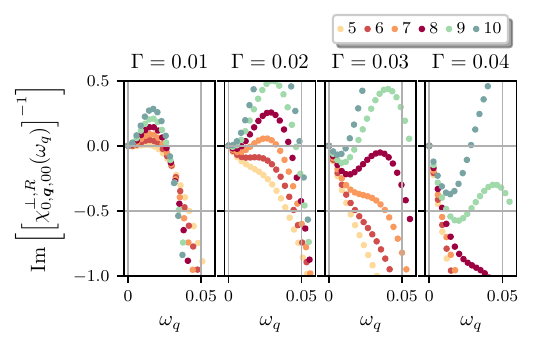}
\caption{Comparison of $\text{Im}\left[\left[\chi^{\bot, R}_{0,\bm{q},00}(\omega_q)\right]^{-1}\right]$ [a.u] for several drive amplitudes $E$ (distinguished by different colors, see legend at top right) and damping strengths $\Gamma$ (shown at the top of each panel) in the small $\omega_q$-regime (near the instability point). 
All the other parameters are the same as in \Cref{fig:Lorenzian_fit} except for $n_{\text{max}}=3$.
}  
\label{fig:chiself_exac}
\end{center}
\end{figure}

Figure~\ref{fig:chiself_exac} presents the systematics of the instability, via plots of $\text{Im}\left[\left[\chi^{\bot, R}_{0,\bm{q},00}(\omega_q)\right]^{-1}\right]\approx -\Gamma_{q}^{\text{eff}}/Z_q$ (see \Cref{eq:fit_chi} ) for different drive strengths and reservoir dampings.  
A positive $\text{Im}\left[\left[\chi^{\bot, R}_{0,\bm{q},00}(\omega_q)\right]^{-1}\right]$ signals an instability. We see that both the drive strength required to induce an instability and the wavevector at which the instability occurs increase with reservoir coupling.

\begin{figure}[b!]
\begin{center} 
\includegraphics[width=0.5\textwidth, angle=-0]{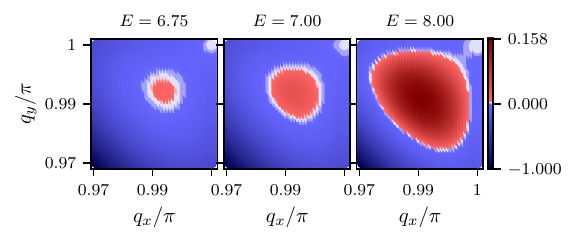}
\caption{
Momentum distribution of $\text{Im}\left[\left[\chi^{\bot, R}_{0,\bm{q},00}(\omega_q)\right]^{-1}\right]$ [a.u] for $\Gamma=0.02$ and several drive amplitudes $E$, which are indicated above the each panel. All the other parameters are the same as in \Cref{fig:Lorenzian_fit}.
}
\label{fig:polediagram}
\end{center}
\end{figure}
Figure ~\ref{fig:polediagram} presents a two dimensional false-color plot of the momentum dependence of $\text{Im}\left[\left[\chi^{\bot, R}_{0,\bm{q},00}(\omega_q)\right]^{-1}\right] \left(\approx -\Gamma_{q}^{\text{eff}}/Z_q\right)$ for different drive strengths for linearly polarized radiation with polarization vector along the zone diagonal. We observe that the region of the instability occurs at momentum $q_x=q_y\neq \pi$. As the drive amplitude $E$ increases, the momentum region of $\text{Im}\left[\left[\chi^{\bot, R}_{0,\bm{q},00}(\omega_q)\right]^{-1}\right]>0$ increases. Changing the polarization changes the momentum region where the instability exists; see \Cref{appendix:CPL}. For smaller interaction regime, larger drive strength or smaller reservoir coupling are required to induce the sign change of $\text{Im}\left[\left[\chi^{\bot, R}_{0,\bm{q},00}(\omega_q)\right]^{-1}\right]$ (see Reference \onlinecite{SM} for additional information).

\section{Origin of the Instability}
\label{sec:_of_the_inetability}
In this section, we provide some mathematical and physical insight into the cause of the sign change of $\text{Im}\left[\left[\chi^{\bot, R}_{0,\bm{q},00}(\omega_q)\right]^{-1}\right]\left(\approx -\Gamma_{q}^{\text{eff}}/Z_q\right)$. Technical details are given in the supplementary material \cite{SM}.

Spin waves in antiferromagnets arise from precession of the fluctuations in the staggered magnetization around fluctuations of the uniform magnetization. In the large $U$ limit of the half filled Hubbard model the precession dynamics and spin wave damping are encoded in the off-diagonal element of \Cref{eq:R_pi_matrix} and we find that the sign change originates from this term $\Pi_{Q,\bm{q}, 00}^{\bot, R}$. In the large $U$ limit the diagonal terms $\Pi_{0,\bm{q},00}^{\bot, R}$ encode the uniform and staggered components of the magnetic susceptibility and we find that the imaginary parts of these terms (which in the large $U$ limit make subleading contributions to the spin wave damping) retain their equilibrium sign and do not exhibit a dynamical instability.

We analyse the Floquet Hamiltonian in the high drive frequency limit, working to leading nontrivial order in the ratio of drive amplitude $E$ to drive frequency $\Omega$. We find that the time averaged components of the polarization matrices are
\begin{widetext}
\begin{align}
  \Pi^{\bot, R}_{\text{0/Q},\bm{q}, 00}\left(\omega\right)&\approx \frac{\mathrm{i} }{4 \pi N} \sum_{\bm{k}}^{\prime} \int d\omega' \mathrm{Tr}\left[G^{R}_{00,\bm{k},\downarrow}\left(\omega^{\prime}\right) (\tau_{0/1})G_{00, \bm{k}+\bm{q},\uparrow}^{K}(\omega^{\prime}-\omega) + G_{00, \bm{k}, \downarrow}^{K}\left(\omega^{\prime}\right) ( \tau_{0/1})  G_{00, \bm{k}+\bm{q}, \uparrow}^{A}(\omega^{\prime}-\omega)  \right], \label{eq:PiR00_block}
\end{align}
\end{widetext}
where the primed sum denotes the summation taken over the magnetic Brillouin zone, and the different $0/Q$ components of the polarization matrix involve different choices $\tau_{0/1}$ of the Pauli matrices. 

The electron propagators $G^{R,A,K}$ are changed from their equilibrium values by virtual transitions into the $n=\pm 1$ Floquet sidebands. These changes may be expressed as contributions to the decay rate of the electron into the reservoir self energies. For our purposes the crucial result is
\begin{equation}
\Im\left(\begin{array}{cc}\Sigma^R_{00,k}(\omega) & \Sigma^K_{00,k}(\omega) \\0 & \Sigma^A_{00,k}(\omega)\end{array}\right)=\Gamma\left(\begin{array}{cc}1+ \frac{2|\epsilon_{\bm{k},1}|^2}{ \Omega^2} & 2 \tanh\left(\frac{\omega}{2T}\right) \\ 0& -1- \frac{2|\epsilon_{\bm{k},1}|^2}{ \Omega^2}\end{array}\right)
\end{equation}
Here, $\epsilon_{\bm{k}, m}=\frac{1}{\Tilde{T}} \int^{\Tilde{T}}_{0} dt e^{\mathrm{i}m\Omega t} \epsilon_{\bm{k}}(t)$ ($\Tilde{T}=2 \pi/\Omega$) and $m\neq 0$ components provide transitions between different  Floquet sidebands. In the presence of a nonequilibrium drive, the fluctuation-dissipation theorem (FDT) is violated in the sense that $\left(\Sigma^R-\Sigma^A\right)\tanh\left(\frac{\omega}{2T}\right)\neq\Sigma^K$.  

An immediate consequence of the FDT breakdown is that the electron  occupation function $n(\omega)=\frac{1}{2}\left(G^R-G^A-G^K\right)$ (momentum and sublattice indices not explicitly written) deviates from the equilibrium value. In equilibrium at $T=0$ with no substrate coupling the lower Hubbard band is fully occupied, the upper Hubbard band is completely empty (mathematically, the $T=0$ equilibrium occupation function is expressed in terms of the electron spectral function $A$ as $\Theta(\omega)A(\omega)$). Spin-wave dissipation arises only from transitions from the lower to the upper Hubbard band; these occur at low frequencies because the coupling to the substrate means that the Hubbard bands (many body density of states) have tails $\sim \Gamma/m_0^2$ overlapping the Fermi energy ($m_0$ is the equilibrium electronic gap in our conventions).

In the nonequilibrium situation considered here the occupancy  of the negative energy states is reduced from the equilibrium value by an amount of order $\frac{|\epsilon_{\bm{k},1}|^2}{ \Omega^2}A(\omega)$ and the occupancy of the positive energy states is greater than zero in a similar way. We may consider this difference as arising from drive induced mixing between the Floquet sidebands. As will be seen immediately, the mixing means that certain transitions, blocked in equilibrium, become allowed. 

With the understanding of the electron propagators in hand, we turn to evaluation of \Cref{eq:PiR00_block}, in the large drive frequency limit $\Omega\gg E_{\bm{k}} \equiv \sqrt{\left(m_0^{(0)}\right)^2+\left(\epsilon_{\bm{k},0}\right)^2}$ ($m_0^{(n)}$ is the Floquet antiferromagnetic mean field along the $z-$axis) as well as expanding \Cref{eq:PiR00_block} up to the second leading order in $\Tilde{t}/\Omega$ and the Floquet off-diagonal block matrix ($\propto J_1\left(E/\Omega\right)$), we obtain (see Reference \onlinecite{SM} for details)
\begin{widetext}
\begin{align}
  \text{Im}\left[\Pi^{\bot, R}_{0,\bm{q}, 00}\left(\omega\right)\right]&\approx\frac{\mathrm{1}}{4 \pi N} \sum_{\bm{k}}^{\prime}\sum_{kl} \int d\omega' B_{\bm{kq},kl,\downarrow} B_{\bm{kq},lk,\downarrow}^{\dagger} M_{\bm{kq}}^{kl} \left(\omega', \omega \right) \label{eq:Impi_0q00}\\
\text{Im}\left[\Pi^{\bot, R}_{Q,\bm{q}, 00}\left(\omega\right)\right]&\approx\frac{\mathrm{1}}{4 \pi N} \sum_{\bm{k}}^{\prime}\sum_{kl} \int d\omega' C_{\bm{kq},kl\downarrow} B_{\bm{kq},lk,\downarrow}^{\dagger} M_{\bm{kq}}^{kl} \left(\omega', \omega \right), \label{eq:Impi_Qq00}
\end{align}
where $B_{\bm{kq},kl,\downarrow}$ and $C_{\bm{kq},kl,\downarrow}$ are matrix products of the unitary matrices which diagonalize the undriven Hamiltonian with the renormalization of the hopping parameter $\Tilde{t} \rightarrow \Tilde{t} J_0(E/\Omega)$ (Magnus limit). The key feature resides in $M_{\bm{kq}}^{kl} \left(\omega', \omega \right)$, which takes the following form in the $T \rightarrow 0$ limit:
\begin{align}
 M_{\bm{kq}}^{kl} \left(\omega', \omega \right) 
   &= \begin{cases} 
   2\Gamma^2 \left( 1 +\frac{|\epsilon_{\bm{k}+\bm{q},1}|^2+|\epsilon_{\bm{k},1}|^2}{\Omega^2}  \right)  \frac{\sgn(\omega')-\sgn(\omega'-\omega)}{\left[{\left(\omega'-\omega-E_{\bm{k}+\bm{q}}^{l}\right)^2+\Gamma^2}\right]\left[\left(\omega'-E_{\bm{k}}^{k}\right)^2+\Gamma^2\right]}>0 & 0<\omega'<\omega \\
   \Gamma^2 \sum_{n=\pm 1}\left(\frac{ |\epsilon_{\bm{k},1}|^2}{\left(\omega'+  n \Omega \right)^2+ \Gamma^2 } - \frac{ |\epsilon_{\bm{k+q},1}|^2}{\left(\omega'-\omega+  n \Omega \right)^2+ \Gamma^2 } \right)\frac{ \left(\sgn(\omega'-\omega)+\sgn(\omega')\right) }{\left[{\left(\omega'-\omega-E_{\bm{k}+\bm{q}}^{l}\right)^2+\Gamma^2}\right]\left[\left(\omega'-E_{\bm{k}}^{k}\right)^2+\Gamma^2\right]} & \omega'<0, \omega'>\omega,
  \end{cases} \label{eq:M_kl_omeprime}
\end{align}
where $l,k$ denote band indices, i.e. $E_{\bm{k}}^{\pm}=\pm E_{\bm{k}}$. The drive-induced breaking of the FDT discussed above makes the second line of \Cref{eq:M_kl_omeprime} nonvanishing.
\end{widetext}

The first line of \Cref{eq:M_kl_omeprime} gives the standard damping term: the difference of sign functions means that one denominator is in the lower, and one in the upper Hubbard bands; at small $\omega$, the  $\omega'$-integration of this term gives $\sim\omega\Gamma^2/E_k^4$ and is only nonvanishing because the coupling to the reservoir broadens the Hubbard bands so they slightly overlap the chemical potential. In the high drive-frequency regime, the driving contribution gives a small correction ($\sim\frac{|\epsilon_{\bm{k},1}|^2}{\Omega^2}\ll 1$) to the non-driving contribution. 

In contrast, the second line is nonzero only out of equilibrium and arises from the drive-induced breaking of the fluctuation-dissipation relation of $G^K$ to the spectral function defined by $G^{R,A}$. The $\omega'$-integration over these regimes allows one to pick up the poles $\omega'=\omega+E_{\bm{k}+\bm{q}}^{l}$ and $\omega'=E_{\bm{k}}^{k}$, corresponding physically to transitions within the upper or within the lower Hubbard bands, which in equilibrium are normally prohibited by the sign function (or Fermi function at non-zero temperature) but are possible here because the virtual excitation to the Floquet sidebands means that the lower Hubbard band states are not completely full and the upper not completely empty.

Multiplying by $B_{\bm{kq},kl,\downarrow}$ and $C_{\bm{kq},kl,\downarrow}$ and carrying out the $\omega'$-integral in \Cref{eq:Impi_Qq00}, we find that the lower case in \Cref{eq:M_kl_omeprime} yields a negative contribution exclusively to $\text{Im}\left[\Pi^{\bot, R}_{Q,\bm{q}, 00}\left(\omega\right)\right]$ (see Reference \onlinecite{SM} for details).

\section{Magnon Distribution Function}
\label{sec:Magnon_distribution_function}
\begin{figure}[t!]
\begin{center} 
\includegraphics[width=0.5\textwidth, angle=-0]{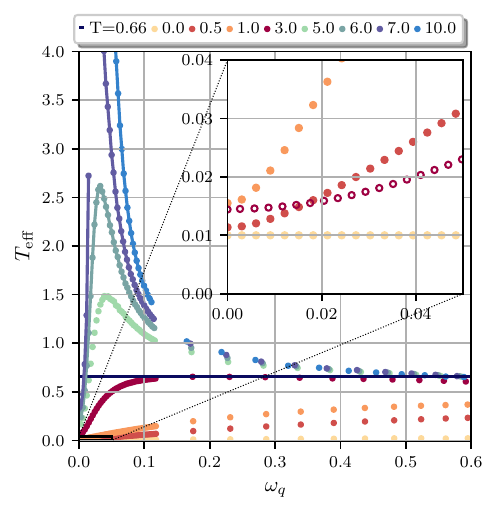}
\caption{Effective temperature defined in \Cref{eq:def_effT} as a function of magnon frequency $\omega_q$ for several drive amplitudes $E$, which are indicated in the legend on top of the plot. In the regime of $\omega_q\sim 0.05$ and $E>5$, we connect plotting markers with lines as a visual guide (The connection is disrupted for $E=7$ and $\omega_q \sim 0.05$ since $\Gamma_{q}^{\text{eff}}<0$). The black line indicates temperature $T=0.66$. All the other parameters are the same as in \Cref{fig:Lorenzian_fit} except for $n_{\text{max}}=3$. In the inset, the $\Gamma=0.2$ case is also plotted as empty circles.
}
\label{fig:3panelplot}
\end{center}
\end{figure}

Because the low energy excitations are very weakly damped (broadening small compared to mode frequency) spin waves, we define the time-averaged magnon distribution function $F_q$ through the ansatz:
\begin{align}
F_q=  \frac{ \chi^{\bot, K}_{0,\bm{q},00}(\omega_q)}{ 2\mathrm{i} \mathrm{Im} \left[\chi^{\bot, R}_{0,\bm{q},00}(\omega_q)\right]} \label{eq:distribution_RPA_deff}
\end{align}
In equilibrium, $F_q=\mathrm{coth}(\omega_q/2T)$. We therefore define an effective temperature in the nonequilibrium case via 
\begin{equation}
F_q=\mathrm{coth}\left(\omega_q/2T_{\text{eff}}\left(\omega_q\right)\right). \label{eq:def_effT}
\end{equation}
In \Cref{fig:3panelplot}, we present the effective temperature $T_{\text{eff}}$ defined in \Cref{eq:def_effT}. In equilibrium, the effective temperature is independent of mode energy and equal to the reservoir temperature. At nonzero, but low drive strength we find, consistent with Reference \onlinecite{Walldorf2019}, that the effective temperature increases with drive strength and with inverse magnon frequency, and that at a particular intermediate drive strength the effective temperature becomes mode frequency independent for a wide range of mode frequencies. However, contrary to Reference \onlinecite{Walldorf2019}, our numerics indicate that where the modes are stable, the effective temperature remains finite and indeed becomes small  as $\delta q\rightarrow 0$, indicating renormalized classical physics at the lowest scales~\cite{Mitra2006}. We note however that numerical resolution issues complicate the analysis of the strongest drive and smallest $\delta q$. Additional analytical insight into the low mode frequency behavior is given in the supplementary material \cite{SM}. As drive strength is increased, a maximum in $T_{\text{eff}}$ emerges at a particular frequency corresponding to the minimum in $\Gamma_{q}^{\text{eff}}$ discussed in the previous section, and as the drive strength reaches a critical value, $T_{\text{eff}}$ diverges at $\delta q=\delta q^\star$. For drive strengths beyond the critical value, the effective temperature is undefined because the $\Gamma_{q}^{\text{eff}}$ becomes negative.

\section{Conclusion}
\label{sec:conclusion}
We use  a Floquet-driven Hubbard system coupled to an electronic reservoir as a model to elucidate the effect of a Floquet drive on electronic collective modes.   We computed the magnon spectrum and distribution function in the antiferromagnetic phase of the model, finding that as the drive strength is increased, a remarkable instability occurs, at which the sign of the magnon damping changes sign at a particular nonzero wave vector leading to an exponential amplification of selected magnon modes and thus to a dynamical instability. Approximate analytical calculations showed that  the instability can be traced back to Floquet-assisted processes active only in the nonequilibrium situation that break the fluctuation-dissipation relation between the Keldysh and retarded/advanced components of the electron propagator and enable the transfer of energy from the reservoir to the magnon system. It is noteworthy that there is no explicit electronic population inversion--the occupancy of the upper Hubbard band remains negligible. The breaking comes because the virtual transition to the Floquet sidebands leads to an additional damping in the retarded/advanced spectral functions, not reflected in the Keldysh component.  It is  important to point out that our assumption of a broadband electron reservoir that can couple both the physical electron bands and the Floquet sidebands is essential. The wavevector of the unstable mode is determined by the interplay of the damping and the frequency, amplitude and polarization of the drive field.

We also revisited the magnon distribution function where  Reference~\onlinecite{Walldorf2019} identified an apparent superthermal divergence of the magnon distribution at zero wavevector, with no sign change of the magnon damping. We show that this interpretation arose from a lack of momentum resolution in the previous calculation, whereby the non-monotonous low-frequency behavior of $T_{\text{eff}}\left(\omega_q\right)$ and eventually the mode instability at $\delta q=\delta q^{\ast}$ for strong enough drive was missed.

The instability uncovered here differs from conventional instabilities driven by nonlinear interactions among excited modes and from drive induced changes in a Hamiltonian parameter that would drive a system through a critical point. The instability has some similarities to the lasing instability, in that a nonequilibrium system becomes susceptible to large-scale stimulated emission of excitations. Two differences are that the stimulated emission implied  by the imaginary part sign change occurs because of an interaction with a reservoir, not because the system is decaying from a highly excited state, and that it  occurs first for modes at a nonzero $\bm q$, unlike the lasing instability which is a property of the long-wavelength limit of the atomic system. The character of the phase emerging beyond the instability point remains an open question. Resolving it will require going beyond Gaussian theory, most likely by formulating an effective field theory of the driven system~\cite{Zelle2024, Daviet2024}. Although this is beyond the scope of this paper, it raises the interesting question of which dynamical patterns~\cite{Cross1993} may arise in Floquet driven dissipative itinerant antiferromagnets. More broadly, our results motivate a search for "stimulated emission" type instabilities in other driven-dissipative Floquet systems. A particularly interesting question is to what extent analogous behavior can arise in ferromagnetic settings (i.e. at $\bm Q=(0,0)$).

\begin{acknowledgments} 
\noindent \textit{Acknowledgments}.---
We acknowledge support from the Max Planck-New York City Center for
Non-Equilibrium Quantum Phenomena. The Flatiron Institute is a division of the Simons Foundation.
T.O acknowledges support from the JSPS Overseas Research Fellowships. MAS was funded by the European Union (ERC, CAVMAT, project no. 101124492). AR was supported by the European Research Council (ERC-2024-SyG- 101167294 ; UnMySt), the Cluster of Excellence Advanced Imaging of Matter (AIM). ME is supported by the Cluster of Excellence ``CUI: Advanced Imaging of Matter'' of the Deutsche Forschungsgemeinschaft (DFG) – EXC 2056 – project ID 390715994. Views and opinions expressed are however those of the author(s) only and do not necessarily reflect those of the European Union or the European Research Council. Neither the European Union nor the European Research Council can be held responsible for them. ME and MAS acknowledge funding by the Deutsche Forschungsgemeinschaft (DFG, German Research Foundation)- 531215165 (Research Unit ‘OPTIMAL’)). 
\end{acknowledgments}

\appendix
\section{Circularly and Elliptically Polarized Light}
\label{appendix:CPL}
\begin{figure}[t!]
\begin{center} 
\includegraphics[width=0.5\textwidth, angle=-0]{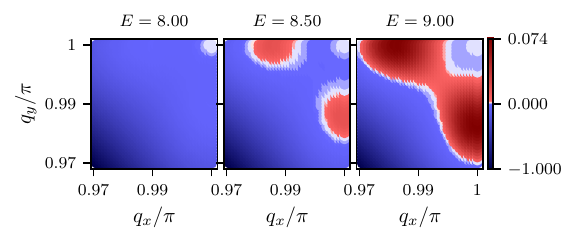}
\includegraphics[width=0.5\textwidth, angle=-0]{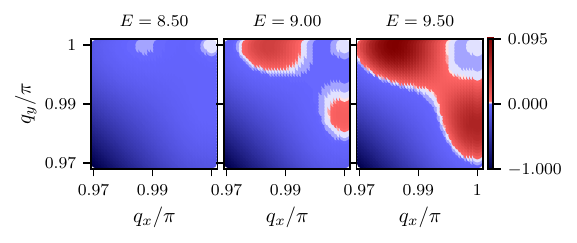}
\includegraphics[width=0.5\textwidth, angle=-0]{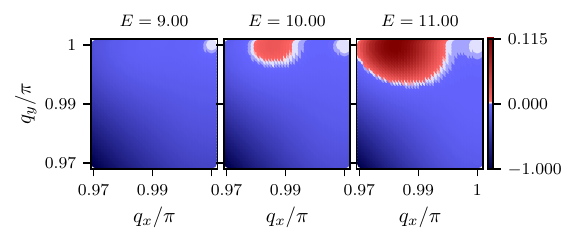}
\caption{
Momentum distribution of $\text{Im}\left[\left[\chi^{\bot, R}_{0,\bm{q},00}(\omega_q)\right]^{-1}\right]$ [a.u] for $\Gamma=0.02$ and several drive amplitudes $E$, which are indicated above each panel. Circularly polarized light is used for upper panel while elliptically polarized light is employed for middle and bottom panels with $(E_x, E_y)=(E,0.9E)$ and $(E_x, E_y)=(E,0.5E)$, respectively. All the other parameters are the same as in \Cref{fig:Lorenzian_fit} except for $n_{\text{max}}=1$.
}
\label{fig:polediagram:CPL}
\end{center}
\end{figure}
Here, we present the results for circularly and elliptically polarized light. The corresponding vector potential may be chosen as $\bm{A}(t)=-\left(\frac{ E_x \mathrm{cos}(\Omega t)}{\Omega},\frac{E_y \mathrm{sin}(\Omega t)}{\Omega}\right)$. The representative results are summarized in \Cref{fig:polediagram:CPL}. The upper panels show the results of circularly polarized light and the first instability appears at a non-zero momentum similar to the case of linearly polarized light found in \Cref{fig:polediagram}. Contrary to the linearly polarized light, the instability momentum is located around either $\bm{q}=(q_x^{\ast}, \pi)$ or $\bm{q}=(\pi, q_y^{\ast})$. The middle panels shows the results of elliptically polarized light with $E_x/E_y=0.9$. Compared to the circularly polarized light presented in the upper panel, now the first instability momentum occurs in $\bm{q}=(q_x^{\ast}, \pi)$, while $\bm{q}=(\pi, q_y^{\ast})$ point also exhibits sign change with reduced magnitude. Finally, the lower panels display results of elliptically polarized light with $E_x/E_y=0.5$. Compared with the $E_x/E_y=0.9$ case, the threshold $E$-value increases, and the instability momentum is found only around $\bm{q}=(q_x^{\ast}, \pi)$ for the range of $E$ considered.

\bibliography{main_Arxiv}%

\end{document}


\title{Supplemental Material: Dynamical instability in a Floquet-Driven Dissipative System}

\author{Takuya Okugawa} 
\email{to2462@columbia.edu}
\affiliation{Department of Physics, Columbia University, New York, NY 10027, USA}
\affiliation{Max Planck Institute for the Structure and Dynamics of Matter, Center for Free Electron Laser Science, 22761 Hamburg, Germany.}

\author{Jens Paaske}
\affiliation{Center for Quantum Devices, Niels Bohr Institute, University of Copenhagen, 2100 Copenhagen, Denmark}

\author{Martin Eckstein}
\affiliation{Institute of Theoretical Physics, University of Hamburg, 20355 Hamburg, Germany}
\affiliation{The Hamburg Centre for Ultrafast Imaging, Hamburg, Germany}

\author{Michael A. Sentef}
\affiliation{Institute for Theoretical Physics and Bremen Center for Computational Materials Science, University of Bremen, 28359 Bremen, Germany}
\affiliation{Max Planck Institute for the Structure and Dynamics of Matter, Center for Free Electron Laser Science, 22761 Hamburg, Germany.}

\author{Angel Rubio}
\affiliation{Max Planck Institute for the Structure and Dynamics of Matter, Center for Free Electron Laser Science, 22761 Hamburg, Germany.}
\affiliation{Center for Computational Quantum Physics, Simons Foundation Flatiron Institute, New York, NY 10010 USA.}
\affiliation{Nano-Bio Spectroscopy Group, Departamento de Fisica de Materiales, Universidad del Pa\'is Vasco, UPV/EHU- 20018 San Sebasti\'an, Spain.}

\author{Andrew J. Millis}
\affiliation{Department of Physics, Columbia University, New York, NY 10027, USA}
\affiliation{Center for Computational Quantum Physics, Simons Foundation Flatiron Institute, New York, NY 10010 USA.}


\pacs{} 
\date{\today} 
\maketitle

\section{Model and method}
\label{sec:model_method}
Here, we show the model Hamiltonian we employed in this work, which was originally used in Reference \onlinecite{Walldorf2019}. Our target system is the half-filled two-dimensional square-lattice Hubbard model with nearest-neighbor hopping in the presence of an applied electromagnetic field with each lattice being connected to a metallic reservoir: 
\begin{align}
    \hat{H}&=\sum_{ij\sigma} h_{ij}(t) \hat{c}^{\dagger}_{i\sigma} \hat{c}_{j\sigma} +  \sum_{i}U\hat{n}_{i\uparrow}\hat{n}_{i\downarrow}+ \hat{H}_{\text{res}} \label{eq:model_Ham} 
\end{align}
where $n_{i\sigma}=c_{i\sigma}^{\dagger}c_{i\sigma}$ and $c_{i\sigma} (c_{i\sigma}^{\dagger})$ is the annihilation (creation) operator for an electron with spin $\sigma$ on the lattice site $i$. $h_{ij}(t)$ is the time-dependent matrix element describing electron hopping from lattice site $j$ to $i$ in the presence of an applied electromagnetic field. We consider a uniform but time-dependent electric field, whose effect is treated by Peierls substitution with a vector potential $A_l(t)$~\cite{Ono2019, Walldorf2019, Aoki2014}. Thus, the non-interacting part of the system Hamiltonian $\sum_{ij\sigma} h_{ij}(t) \hat{c}^{\dagger}_{i\sigma} \hat{c}_{j\sigma}$ is diagonal in momentum space. Thus, in units where $\hbar=e=a=1$, the electron dispersion is given by 
\begin{align}
    \epsilon_{\bm{k}}(t)&=-2\Tilde{t} \sum_{l=x,y}\mathrm{cos}\left(k_l+  A_l(t) \right) \label{equantion:e_km},
\end{align}
where $\Tilde{t}$ is the hopping parameter, supplied by a tilde to distinguish it from the time variable $t$. As in the main text, we express all energies in units of $\Tilde{t}$. 
In the following, we consider the linearly polarized light for the main text and supplemental material, and the circularly or elliptically polarized light for the appendix. The former is given as
\begin{equation}
A_l(t) =-\frac{E_l \mathrm{sin}(\Omega t)}{\Omega} 
\label{eq:A_linear},   
\end{equation}
while the latter is 
\begin{equation}
A(t)=-\left(\frac{ E_x \mathrm{cos}(\Omega t)}{\Omega},\frac{E_y \mathrm{sin}(\Omega t)}{\Omega}\right), 
\label{eq:A_circular}   
\end{equation}
$\hat{H}_{\text{res}}$ describes a weak tunnel coupling to an infinite-bandwidth reservoir with a flat density of states, giving rise to a constant inverse electron lifetime $\Gamma$.  

We set the chemical potential of all the reservoirs to be identically zero to ensure the half-filling ($\mu=0$). We assume that all reservoirs are equilibrated with the same temperature ($\beta=1/T$) and each reservoir couples to the single associated lattice site with the same coupling strength ($\Gamma$), where $T$ is the bath temperature and the Boltzmann constant is set to be $k_B=1$.   

As in Reference \onlinecite{Walldorf2019}, we decouple the interaction term $\sum_{i}U\hat{n}_{i\uparrow}\hat{n}_{i\downarrow}$ via a magnetic-channel using Hubbard-Stratonovich transformation~\cite{altland_simons_2010,Kamenev_2023} and introduce the magnetization field $\bm{m}_0(t)$. 
We choose the direction of the magnetization field along the $z-$axis and focus on the antiferromagnetic phase: $\bm{m}_0(t) = m_0(t) \hat{z} e^{i \bm{Q} \cdot \bm{R}_i}$, where $\bm{R}_i$ is the lattice vector and $\bm{Q}=\left(\pi, \pi \right)$. Then, we again write the antiferromagnetic mean-field equation initially introduced in Reference \onlinecite{Walldorf2019}, which reads
\begin{align}
m_{0}^{(m)}&=\frac{I}{2N\mathrm{i}} \sum'_{\bm{k}} \int_{-\infty}^{\infty} \frac{d\omega}{2\pi} \text{Tr}  \left[  G_{\bm{k},m0}^{K} (\omega) \tau_1 \otimes \sigma_z  \right],  \label{equation:AFM_MF:m_I}      
\end{align}
where $G$ is the mean-field Floquet Green’s function which has a matrix structure in (momentum-)spinor ($\tau$), spin ($\sigma$), and Floquet space. The primed sum denotes the summation taken over the magnetic Brillouin zone. The Keldysh Green's function and the self-energy in Floquet space can be expressed as
\begin{align}
G_{\bm{k},mn}^{K}(\omega) &= \sum_{m'n'} G^{R}_{\bm{k},mm'}(\omega)\Sigma^{K}_{m'n'}(\omega)G^{A}_{\bm{k},n'n}(\omega),\\
\Sigma^{K}_{mn}(\omega) &= -2 \mathrm{i} \Gamma \left(\tanh{ \frac{\omega+n\Omega}{2 T} } \right)  \delta_{mn} \tau_0 \otimes \sigma_0, 
\end{align}
where the indices ($m,n$) correspond to the Floquet space indices.
From the Dyson equation, the free Green's function, and the reservoir self-energy for the retarded/advanced components in Floquet space, one can find the inverse of the reservoir-dressed retarded/advanced Green's function~\cite{Walldorf2019, Walldorf_thesis, Aoki2014}:  
\begin{align}
 G_{\bm{k},mn}^{R/A -1}(\omega) &= (\omega+n \Omega \pm \mathrm{i}\Gamma) \delta_{mn} \tau_0 \otimes \sigma_0 -h_{\bm{k},mn}  \\
h_{\bm{k},mn}&=\epsilon_{\bm{k}, m-n} \tau_3 \otimes \sigma_0 - m_0^{(m-n)} \tau_1 \otimes \sigma_3    \label{eq:equiv_H}
\end{align}
where 
\begin{align} 
\epsilon_{\bm{k}, m}=\frac{1}{\Tilde{T}} \int^{\Tilde{T}}_{0} dt e^{\mathrm{i}m\Omega t} \epsilon_{\bm{k}}(t).
\end{align}
where $\Tilde{T}=2 \pi/\Omega$.

By following the same procedure as in Reference \onlinecite{Walldorf2019}, we consider the fluctuation field with respect to the mean-field found from \Cref{equation:AFM_MF:m_I}. They are dictated by the Retarded/Advanced/Keldysh electron Green's function bubbles, and their Floquet space representations are found as follows:
\begin{align}
\Pi^{\mu\nu, R}_{0/Q,\bm{q}, mn}(\omega)&=-\frac{1}{4 \pi \mathrm{i} N} \sum_{m', \bm{k}} \int d\omega' \mathrm{Tr} \left[(\tau_{0} \otimes \sigma_{\mu}) G^{R}_{mm'}(\omega') (\tau_{0/1} \otimes \sigma_{\nu})G_{m'n, \bm{k}+\bm{q}}^{K}(\omega'-(\omega+n\Omega)) \right. \notag\\
&\left. + ( \tau_{0} \otimes \sigma_{\mu} )G_{mm', \bm{k}}^{K}(\omega') ( \tau_{0/1} \otimes \sigma_{\nu} )   G_{m'n, \bm{k}+\bm{q}}^{A}(\omega'-(\omega+n\Omega))   \right]    \label{eq:Retarded_bubble} \\
\Pi^{\mu\nu, A}_{0/Q,\bm{q}, mn}(\omega)&=-\frac{1}{4 \pi \mathrm{i} N} \sum_{m', \bm{k}} \int d\omega' \mathrm{Tr} \left[ ( \tau_{0} \otimes \sigma_{\mu} )G_{mm', \bm{k}}^{A}(\omega') ( \tau_{0/1} \otimes \sigma_{\nu} )   G_{m'n, \bm{k}+\bm{q}}^{K}(\omega'-(\omega+n\Omega))\right. \notag\\
&\left. +  (\tau_{0} \otimes \sigma_{\mu}) G^{K}_{mm'}(\omega') (\tau_{0/1} \otimes \sigma_{\nu})G_{m'n, \bm{k}+\bm{q}}^{R}(\omega'-(\omega+n\Omega))  \right]  \\
\Pi^{\mu\nu, K}_{0/Q,\bm{q}, mn}(\omega)&=-\frac{1}{4 \pi \mathrm{i} N} \sum_{m', \bm{k}} \int d\omega' \mathrm{Tr} \left[(\tau_{0} \otimes \sigma_{\mu}) G^{A}_{mm'}(\omega') (\tau_{0/1} \otimes \sigma_{\nu})G_{m'n, \bm{k}+\bm{q}}^{R}(\omega'-(\omega+n\Omega)) \right. \notag\\
&\left. + ( \tau_{0} \otimes \sigma_{\mu} )G_{mm', \bm{k}}^{R}(\omega') ( \tau_{0/1} \otimes \sigma_{\nu} )   G_{m'n, \bm{k}+\bm{q}}^{A}(\omega'-(\omega+n\Omega))   \right. \notag\\
&\left. + ( \tau_{0} \otimes \sigma_{\mu} )G_{mm', \bm{k}}^{K}(\omega') ( \tau_{0/1} \otimes \sigma_{\nu} )   G_{m'n, \bm{k}+\bm{q}}^{K}(\omega'-(\omega+n\Omega)) \right] \label{eq:Keldysh_bubble}
\end{align}
where $\sigma_{\pm}=\frac{\sigma_x\pm\mathrm{i}\sigma_y}{2}$. 
The Retarded/Advanced/Keldysh transverse fluctuation matrix propagators are then given as
\begin{align}
\bm{\chi}^{\bot, R}_{\bm{q},mn}(\omega) & \equiv \left[\frac{1}{2I}-\bm{\Pi}^{\bot, R}_{\bm{q}} (\omega)  \right]_{mn}^{-1} \label{eq:chiR_original}\\
\bm{\chi}^{\bot, K}_{\bm{q},mn}(\omega) & \equiv \Biggl( \left[\frac{1}{2I}-\bm{\Pi}^{\bot, R}_{\bm{q}} (\omega)  \right]^{-1}   \bm{\Pi}^{\bot, K}_{\bm{q}} (\omega)  \left[\frac{1}{2I}-\bm{\Pi}^{\bot, A}_{\bm{q}} (\omega)  \right]^{-1}      \Biggr)_{mn}  \label{eq:chiK_original}
\end{align}
where the bold character corresponds to the matrix structure consisting of spinor and Floquet degree of freedom. The spinor matrix components are defined as follows:
\begin{equation}
\bm{\Pi}^{\bot, R}_{\bm{q},mn}=
\begin{pmatrix}
\Pi^{\bot, R}_{0,\bm{q},mn} & \Pi^{\bot, R}_{Q,\bm{q},mn} \\
\Pi^{\bot, R}_{Q,\bm{q},mn} & \Pi^{\bot, R}_{0, \bm{Q}+\bm{q},mn} \\
\end{pmatrix}. \label{eq:R_pi_matrix}
\end{equation}
Finally, we define a time-averaged distribution function, $F$, by the ansatz:
\begin{align}
    \chi^{\bot, K}_{0,\bm{q},00}(\omega) &= 2\mathrm{i} \mathrm{Im} \left[\chi^{\bot, R}_{0,\bm{q},00}(\omega)\right] F(\bm{q},\omega) \approx 2\mathrm{i}Z_q \delta(|\omega|-\omega_q)F_q, \label{eq:distribution_RPA_deff}
\end{align}
where $F_q=F(\bm{q},\omega_q)$, referring only to the mode energy $\omega_q$. In equilibrium, $F_q=\mathrm{coth}(\omega_q/2T)$.  
Thus, we can find the time-averaged distribution function and the effective temperature $T_{\text{eff}}$ can be determined as follows:
\begin{align}
   F_q &=  \frac{\chi^{\bot, K}_{0,\bm{q},00}(\omega_q)}{2\mathrm{i} \mathrm{Im} \left[\chi^{\bot, R}_{0,\bm{q},00}(\omega_q)\right]} \label{eq:distribution_RPA_numerical}, \\
   F_q &= \mathrm{coth}(\omega_q/2T_{\text{eff}}\left(\omega_q\right)).
\end{align}

\section{Result and discussion}
\label{sec:Result_discussion}
\begin{figure}[t!]
    \centering      
     \begin{minipage}[t]{0.45\textwidth}
        \includegraphics[width=\textwidth, angle=-0]{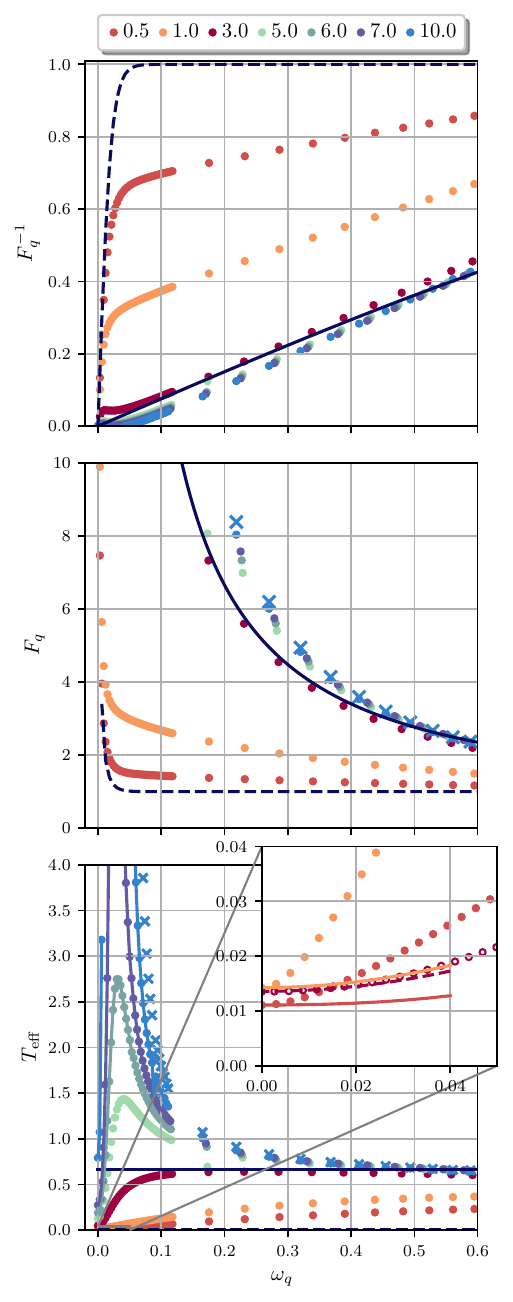}
        \caption{Upper, middle, and lower panels show the inverse of the magnon-distribution $F_q^{-1}$, the magnon-distribution $F_q$, and the effective temperature $T_{\text{eff}}$ as a function of magnon frequency $\omega_q$ for several drive amplitudes $E$, which are indicated in the legend on top of the upper panel. $F_q$ is computed from \Cref{eq:2ndF_q}. Crossed markers in the two lower panels indicate the results computed from \Cref{eq:dKdR}. Full lines in the inset show the results of  \Cref{eq:Fq_wzero_limit}, respectively. The parameters are $I=5$, $T=0.01$, $\Omega=30$, and $\Gamma=0.02$. In the inset, the $\Gamma=0.2$ case is also plotted as empty circles. Solid and dashed black curve denote the equilibrium distribution function for $T=0.66$ and $T=0.01$, respectively.}
\label{fig:expand_ana}
    \end{minipage}
    \hspace{0.1cm}
    \begin{minipage}[t]{0.45\textwidth}
        \includegraphics[width=\textwidth, angle=-0]{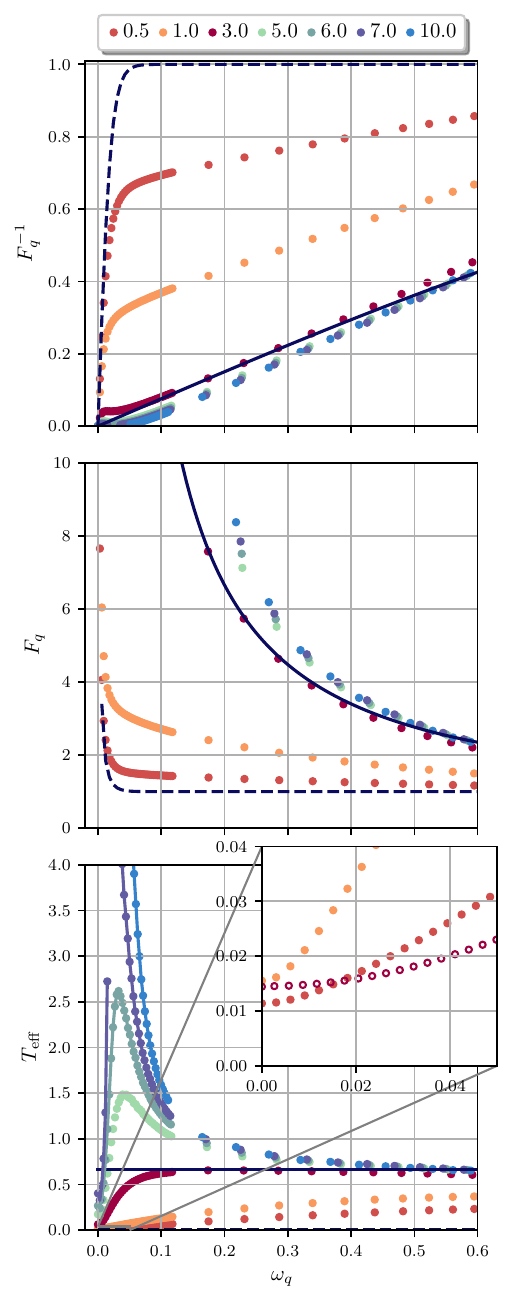}
        \caption{The same plots as \Cref{fig:expand_ana}. The magnon-distribution $F_q$ is computed from \Cref{eq:distribution_RPA_deff}.}
        \label{fig:full_3panel_Fq}
    \end{minipage}      
\end{figure}      

\begin{figure}[t!]
    \centering      
    \begin{minipage}[t]{0.45\textwidth}
        \includegraphics[width=\textwidth, angle=-0]{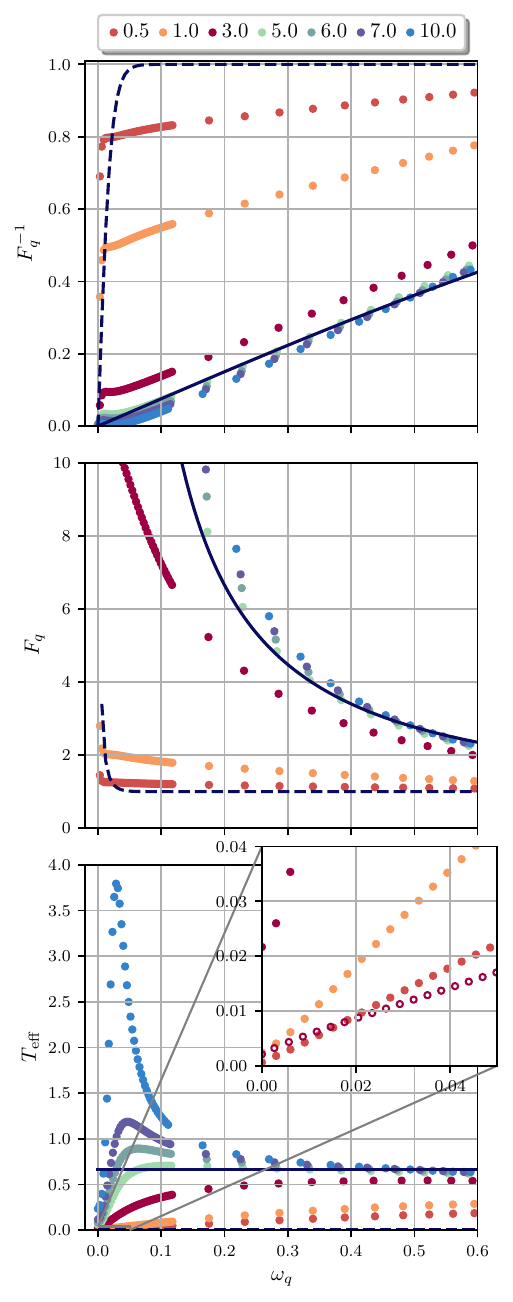}
        \caption{The same plots as \Cref{fig:expand_ana} except that $\delta \Pi^{\bot,R/K}$ functions used to compute $F_q$ are expanded up to the second order in $1/\Omega$ as performed in \Cref{subsubsec:m1omega_expansion}.}
        \label{fig:expandw2_ana}
    \end{minipage}      
\end{figure}      

In this section, we present quasi-analytic results~\cite{herre2025} to get basic understanding of the feature of $F_q$. First, we confirmed that all features of $F_q$ can be also found with smaller Floquet cutoff $n_{\text{max}}=1$ whereas $n_{\text{max}}=3$ is used in Fig. 1 of the main text and Ref.~\onlinecite{Walldorf2019}. Furthermore, we found that taking simply the time-averaged components of the polarization bubble function $\Pi^{R/A/K}$ in the spinor matrix introduced in \Cref{eq:R_pi_matrix} for the respective transverse fluctuation matrix $\chi^{R/A/K}$ in \Cref{eq:chiR_original,eq:chiK_original} is well approximated by:
\begin{align}
\bm{\chi}^{\bot, R}_{\bm{q},00}(\omega) & \approx \left[\frac{1}{2I}-\bm{\Pi}^{\bot, R}_{\bm{q}, 00} (\omega)  \right]^{-1} \label{eq:XaiR00=PiR00_inv} \\
\bm{\chi}^{\bot, K}_{\bm{q},00}(\omega) & \approx \Biggl( \left[\frac{1}{2I}-\bm{\Pi}^{\bot, R}_{\bm{q}} (\omega)  \right]^{-1}_{00}   \bm{\Pi}^{\bot, K}_{\bm{q},00} (\omega)  \left[\frac{1}{2I}-\bm{\Pi}^{\bot, A}_{\bm{q}} (\omega)  \right]^{-1}_{00}      \Biggr) \notag\\
& \approx \Biggl( \left[\frac{1}{2I}-\bm{\Pi}^{\bot, R}_{\bm{q}, 00} (\omega)  \right]^{-1}   \bm{\Pi}^{\bot, K}_{\bm{q},00} (\omega)  \left[\frac{1}{2I}-\bm{\Pi}^{\bot, A}_{\bm{q},00} (\omega)  \right]^{-1}      \Biggr)  \label{eq:XaiK00=XaiR00_XaiA00} \\
\bm{\Pi}^{\bot, R}_{\bm{q}, 00} (\omega) & \approx \bm{\Pi}^{\bot, A, \dagger}_{\bm{q}, 00} (\omega). 
\end{align}
Moreover, we further express the above time-averaged polarizability function $\Pi^{R/K}$ introduced in \Cref{eq:Retarded_bubble,eq:Keldysh_bubble} by the time-averaged Retarded/Advanced/Keldysh-Green's functions:
\begin{align}
\Pi^{\bot, R}_{0/Q,\bm{q}, 00}\left(\omega\right) &\approx -\frac{1}{4 \pi \mathrm{i} N} \sum_{\bm{k}} \int d\omega' \mathrm{Tr} \left[(\tau_{0} \otimes \sigma_{+}) G^{R}_{00,\bm{k}}\left(\omega^{\prime}\right) (\tau_{0/1} \otimes \sigma_{-})G_{00, \bm{k}+\bm{q}}^{K}(\omega^{\prime}-\omega) \right. \notag\\
&\left. + ( \tau_{0} \otimes \sigma_{+} )G_{00, \bm{k}}^{K}\left(\omega^{\prime}\right) ( \tau_{0/1} \otimes \sigma_{-} )   G_{00, \bm{k}+\bm{q}}^{A}(\omega^{\prime}-\omega)   \right] \label{eq:time_ave_R} \\
\Pi^{\bot, K}_{0/Q,\bm{q}, 00}\left(\omega\right)&\approx-\frac{1}{4 \pi \mathrm{i} N} \sum_{\bm{k}} \int d\omega' \mathrm{Tr} \left[(\tau_{0} \otimes \sigma_{+}) G^{A}_{00, \bm{k}}\left(\omega^{\prime}\right) (\tau_{0/1} \otimes \sigma_{-})G_{00, \bm{k}+\bm{q}}^{R}(\omega^{\prime}-\omega) \right. \notag\\
&\left. + ( \tau_{0} \otimes \sigma_{+} )G_{00, \bm{k}}^{R}\left(\omega^{\prime}\right) ( \tau_{0/1} \otimes \sigma_{-} )   G_{00, \bm{k}+\bm{q}}^{A}(\omega^{\prime}-\omega)   \right. \notag\\
&\left. + ( \tau_{0} \otimes \sigma_{+} )G_{00, \bm{k}}^{K}\left(\omega^{\prime}\right) ( \tau_{0/1} \otimes \sigma_{-} )   G_{00, \bm{k}+\bm{q}}^{K}(\omega^{\prime}-\omega) \right] \label{eq:time_ave_K} 
\end{align}
Thus, our aim is to find the time-averaged Retarded/Advanced/Keldysh-Green's functions with Floquet cutoff $n_{\text{max}}=1$.

Since each block Hamiltonian given in \Cref{eq:equiv_H} is diagonal in the spin space, we can simplify \Cref{eq:time_ave_R}. 
Defining "up" and "down" Green's functions, we get    
\begin{align}
\Pi^{+-, R}_{0/Q,\bm{q}, 0}\left(\omega\right)&=-\frac{1}{4 \pi \mathrm{i} N} \sum_{\bm{k}} \int d\omega' \mathrm{Tr} \left[G^{R}_{00,\bm{k},\downarrow}\left(\omega^{\prime}\right) (\tau_{0/1})G_{00, \bm{k}+\bm{q},\uparrow}^{K}(\omega^{\prime}-\omega) \right. \notag \\
&\left. + G_{00, \bm{k}, \downarrow}^{K}\left(\omega^{\prime}\right) ( \tau_{0/1})   G_{00, \bm{k}+\bm{q}, \uparrow}^{A}(\omega^{\prime}-\omega)   \right]    \label{eq:Retarded_bubble_block}
\end{align}
One can do the same operation for the Keldysh polarization bubble in \Cref{eq:time_ave_K}.  
The retarded Green's function we want to obtain is given as
\begin{align}
G^{R}_{mn,\bm{k},\sigma}&=\left[  M_{\bm{k},\sigma} \right]^{-1}_{mn},  \label{eq:GR_tridigonal}
\end{align}
The matrix $M$ is block-tridiagonal (block consists of the spinor degree of freedom) in Floquet space. The block diagonal $M_{nn}$ and non-diagonal $M_{mn}$ components are given as follows:
\begin{align}
M_{nn,\bm{k},\sigma} &= (\omega+n \Omega + \mathrm{i} \Gamma) \tau_0 - \Tilde{h}_{\bm{k},\sigma}  \\
M_{mn,\bm{k},\sigma} &= -\epsilon_{\bm{k},m-n}\tau_3  \\
\Tilde{h}_{\bm{k},\sigma} &=  \epsilon_{\bm{k},0} (\tau_3) -  \sigma m^{(0)}_{0} (\tau_1  ), 
\end{align}
where we ignored $m^{(\pm1)}_{0}$ since $m^{(\pm1)}_{0}/m^{(0)}_{0}\ll 1$. Note that $M_{mn}$ and $\Tilde{h}$ have matrix structure in spinor degree of freedom. We can take the inverse of the $M$ matrix by employing the inversion scheme for the block-tridiagonal matrix~\cite{Godfrin_1991, Eissing2016}. Furthermore, we expand $G^{R}_{mn,\bm{k},\sigma}$ up to second order in the off-diagonal Floquet component $\epsilon_{\bm{k},1}\tau_3$, which is proportional to $J_1(\frac{E}{\Omega})$ and $J_0(\frac{E}{\Omega})\gg J_1(\frac{E}{\Omega})$ in limit of $E/\Omega\ll1$. The lowest order corresponds to the Magnus limit, which simply ends up with the equilibrium magnon distribution function, $F_q=\coth{\frac{\beta \omega}{2}}$ due to the fluctuation-dissipation theorem. The next lowest order is the second order, which we present here.

Defining 
\begin{align}
   \alpha_n &=M_{nn}, \\
   \beta_n &=\tau_z M_{nn}^{-1} \tau_z, \\
   x &=  \left(\frac{|\epsilon_{\bm{k}, 1}|}{\Tilde{t}}\right)^2 \propto J_1(\frac{E}{\Omega})^2, \\
   R_n &=  -2 \mathrm{i} \Gamma \tanh{\left(\frac{ \omega+ n \Omega }{2T}\right)},
\end{align}
and considering up to the second order in $\epsilon_{\bm{k}, 1}/\Tilde{t}$ (first order in $x$), one obtains
\begin{align}
     G^{R}_{00}& \approx \alpha_{0}^{-1} +   x \Tilde{t}^2\alpha_{0}^{-1} \left(\beta_{-1}+\beta_{1}  \right) \alpha_{0}^{-1} \label{eq:GR_tridigonal_expand_Ew}\\
     G^{K}_{00}  &\approx R_0 \left|  \alpha^{-1}_0 \right|^2 + x \Tilde{t}^2\left\{ R_0 2\text{Re}    \left[\alpha_{0}^{-1}  \alpha_{0}^{-1, \ast} \left(\beta_{-1}^{\ast}+\beta_{1}^{\ast}  \right) \alpha_{0}^{-1, \ast}\right] + R_{-1} \left|  \alpha^{-1}_0 \beta_{-1}   \right|^2+   R_{1} \left|  \alpha^{-1}_0 \beta_{1}   \right|^2  \right\} \label{eq:GK_tridigonal_expand_Ew}
\end{align}
where we suppressed the momentum indices $\bm{k}$ and frequency dependence $\omega$. Note that \Cref{eq:GR_tridigonal_expand_Ew,eq:GK_tridigonal_expand_Ew} and $\alpha_n$, $\beta_n$ have matrix structure in spinor degree of freedom.   
We also used the following definitions $|A|^2 \equiv A A^{\dagger}$ and $2\mathrm{Re}[A] \equiv A +A^{\dagger}$, where $A$ is a matrix consisting of $\alpha_0, \beta_{-1/1}$. Upon inserting \Cref{eq:GR_tridigonal_expand_Ew,eq:GK_tridigonal_expand_Ew} into \Cref{eq:Retarded_bubble_block} and the corresponding Keldysh component, and keeping terms up tp second order in $\epsilon_{\bm{k}, 1}/\Tilde{t}$ (first order in $x$), one obtains, after some algebra, the final retarded polarization function $\Pi^{\bot, R/K}_{0,\bm{q}, 00}(\omega)$ as follows:

\begin{align}
\Pi^{\bot, R}_{0,\bm{q}, 00}(\omega) &=\Pi^{\bot, R}_{\text{mag},0,\bm{q}}(\omega)  +  \delta\Pi^{\bot,R}_{0, \bm{q}}(\omega)  \label{eq:PiR_sep}\\
\delta\Pi^{\bot,R}_{0, \bm{q}}(\omega)  &\equiv \frac{\mathrm{i}}{4 \pi  N} \sum_{\bm{k}} \left[|\epsilon_{\bm{k},1}|^2  d\Pi^{R}_{\bm{k},\bm{q}, \downarrow}(\omega) -|\epsilon_{\bm{k}+\bm{q},1}|^2  d\Pi^{R,\ast}_{\bm{k}+\bm{q},-\bm{q},\uparrow}(-\omega) \right]\\
   d\Pi^{R}_{\bm{k},\bm{q},\sigma}(\omega)& \equiv \int d\omega' \mathrm{Tr} \left[ R_{\bm{k}\bm{q}\sigma}^{\bot}(\omega^{\prime}, \omega) \right]  \\
   \mathrm{Tr} \left[ R_{\bm{k}\bm{q}\sigma}^{\bot}(\omega^{\prime}, \omega) \right] &\equiv \sum_{ijkl} A_{ji, \bm{k}\sigma}A_{ik, \bm{k}\sigma}B_{kl, \bm{k q}\sigma}B_{jl, \bm{k q}\sigma} M^{R}_{jikl,\bm{k q}}(\omega^{\prime}, \omega) \label{eq:R_tensor_sum}\\
   M^{R}_{jikl,\bm{k q}}(\omega^{\prime}, \omega)&\equiv\left[ 
   R_{0, \bm{k}+\bm{q}} d_{0,\bm{k}}^{j} (d_{-1,\bm{k}}^{i}+d_{1,\bm{k}}^{i})d_{0,\bm{k}}^{k}d_{0,\bm{k}+\bm{q}}^{l}d_{0,\bm{k}+\bm{q}}^{l,\ast} \right. \notag\\   
  &+ R_{0, \bm{k}} d_{0,\bm{k}}^{j} (d_{-1,\bm{k}}^{i}+d_{1,\bm{k}}^{i})d_{0,\bm{k}}^{k}d_{0,\bm{k}}^{k,\ast} d_{0,\bm{k}+\bm{q}}^{l,\ast} \notag\\
  &+ R_{0, \bm{k}} d_{0,\bm{k}}^{j} d_{0,\bm{k}}^{j,\ast} (d_{-1,\bm{k}}^{i,\ast}+d_{1,\bm{k}}^{i,\ast})d_{0,\bm{k}}^{k,\ast} d_{0,\bm{k}+\bm{q}}^{l,\ast} \notag\\
  &+R_{-1, \bm{k}} d_{0,\bm{k}}^{j} (d_{-1,\bm{k}}^{i}d_{-1,\bm{k}}^{i,\ast}) d_{0,\bm{k}}^{k,\ast} d_{0,\bm{k}+\bm{q}}^{l,\ast} \notag\\
  &+\left.R_{1, \bm{k}} d_{0,\bm{k}}^{j} (d_{1,\bm{k}}^{i}d_{1,\bm{k}}^{i,\ast}) d_{0,\bm{k}}^{k,\ast} d_{0,\bm{k}+\bm{q}}^{l,\ast}    \right],
\end{align}
where $\Pi^{\bot, R}_{\text{mag},0,\bm{q}}(\omega)$ is the undriven contribution with the renormalization of the hopping parameter $\Tilde{t} \rightarrow \Tilde{t} J_0(E/\Omega)$ (Magnus limit), which has already been well-studied~\cite{Walldorf2019, Walldorf_thesis}.   
The Keldysh component reads 
\begin{align}
\Pi^{\bot, K}_{0,\bm{q}, 00}(\omega) &=\Pi^{\bot, K}_{\text{mag},0,\bm{q}}(\omega)  +  \delta\Pi^{\bot,K}_{0, \bm{q}}(\omega)  \label{eq:PiK_sep}\\
\Pi^{\bot, K}_{\text{mag},0,\bm{q}}(\omega) &=  \mathrm{coth}\frac{\beta}{2}\left(\omega\right) \Bigl(2 \mathrm{i} \text{Im} \left[  \Pi^{\bot, R}_{\text{mag},0,\bm{q}}(\omega)   \right] \Bigr)         \\
\delta\Pi^{\bot,K}_{0, \bm{q}}(\omega)  &\equiv \frac{\mathrm{i}}{4 \pi  N} \sum_{\bm{k}} \left[|\epsilon_{\bm{k},1}|^2  d\Pi^{K}_{\bm{k},\bm{q}, \downarrow}(\omega) +|\epsilon_{\bm{k}+\bm{q},1}|^2  d\Pi^{K}_{\bm{k}+\bm{q},-\bm{q},\uparrow}(-\omega) \right]\\
d\Pi^{K}_{\bm{k},\bm{q},\sigma}(\omega)& \equiv \int d\omega' \mathrm{Tr} \left[ K_{\bm{k}\bm{q}\sigma}^{\bot}(\omega^{\prime}, \omega) \right]  \\
   \mathrm{Tr} \left[ K_{\bm{k}\bm{q}\sigma}^{\bot}(\omega^{\prime}, \omega) \right] &\equiv \sum_{ijkl} A_{ji, \bm{k}\sigma}A_{ik, \bm{k}\sigma}B_{kl, \bm{k}\bm{q}\sigma}B_{jl, \bm{k}\bm{q}\sigma} M^{K}_{jikl,\bm{k}\bm{q}}(\omega^{\prime}, \omega) \label{eq:K_tensor_sum}\\
   M^{K}_{jikl,\bm{k}\bm{q}}(\omega^{\prime}, \omega)&\equiv\left[R_{22, \bm{k}} d_{0,\bm{k}}^{j} (d_{-1,\bm{k}}^{i}+d_{1,\bm{k}}^{i})d_{0,\bm{k}}^{k}d_{0,\bm{k}}^{k,\ast} d_{0,\bm{k}+\bm{q}}^{l} d_{0,\bm{k}+\bm{q}}^{l,\ast}  \right. \notag\\   
  &+ R_{22, \bm{k}} d_{0,\bm{k}}^{j} d_{0,\bm{k}}^{j,\ast} (d_{-1,\bm{k}}^{i,\ast}+d_{1,\bm{k}}^{i,\ast})d_{0,\bm{k}}^{k,\ast} d_{0,\bm{k}+\bm{q}}^{l}d_{0,\bm{k}+\bm{q}}^{l,\ast} \notag\\
  &+R_{-12, \bm{k}} d_{0,\bm{k}}^{j} (d_{-1,\bm{k}}^{i}d_{-1,\bm{k}}^{i,\ast}) d_{0,\bm{k}}^{k,\ast} d_{0,\bm{k}+\bm{q}}^{l}d_{0,\bm{k}+\bm{q}}^{l,\ast} \notag\\
  &+\left.R_{12, \bm{k}} d_{0,\bm{k}}^{j} (d_{1,\bm{k}}^{i}d_{1,\bm{k}}^{i,\ast}) d_{0,\bm{k}}^{k,\ast} d_{0,\bm{k}+\bm{q}}^{l}d_{0,\bm{k}+\bm{q}}^{l,\ast}    \right] \label{eq:2ndkeldysh_last}.
\end{align}
The quantities used from \Cref{eq:PiR_sep} to \Cref{eq:2ndkeldysh_last} are as follows:
\begin{align}
&R_{n, \bm{k}} = -2 \mathrm{i} \Gamma \tanh{\frac{\beta}{2}\left( \omega^{\prime}+ n \Omega \right)}\\
&R_{n, \bm{k}+\bm{q}} = -2 \mathrm{i} \Gamma \tanh{\frac{\beta}{2}\left( \omega^{\prime}-\omega + n \Omega \right)}\\
  &A_{\bm{k},\sigma} \equiv U_{\bm{k},\sigma}^{\dagger} \tau_3 U_{\bm{k},\sigma} \label{eq:start:A}\\ 
  &B_{\bm{kq},\sigma} \equiv U_{\bm{k},\sigma}^{\dagger} U_{\bm{k}+\bm{q},\bar{\sigma}} = B_{\bm{k}+\bm{q},-q,\Bar{\sigma}}^{\dagger}\\
  &C_{\bm{kq},\sigma} \equiv U_{\bm{k},\sigma}^{\dagger} \tau_{1} U_{\bm{k}+\bm{q},\bar{\sigma}} = C_{\bm{k}+\bm{q},-q,\Bar{\sigma}}^{\dagger}\\
  &U_{\bm{k},\sigma}^{\dagger}  h_{\bm{k},\sigma} U_{\bm{k},\sigma} = E_{\bm{k}} \tau_3 \\
  &h_{\bm{k},\sigma}=\epsilon_{\bm{k},0} \tau_3 \mp m^{(0)} \tau_1\\
  &\alpha^{-1}_{nn,\bm{k}\sigma}=  \left(U_{\bm{k},\sigma} U_{\bm{k},\sigma}^{\dagger} M_{nn,\bm{k}\sigma} U_{\bm{k},\sigma} U_{\bm{k},\sigma}^{\dagger} \right)^{-1} =U_{\bm{k},\sigma} d_{n,\bm{k}}  U_{\bm{k},\sigma}^{\dagger} \\
  &d_{n,\bm{k}}^{\pm}=\left(\omega^{\prime}+n \Omega + \mathrm{i} \Gamma \mp E_{\bm{k}} \right)^{-1} \\
  &d_{n,\bm{k}+\bm{q}}^{\pm}=\left(\omega^{\prime}-\omega+n \Omega + \mathrm{i} \Gamma \mp E_{\bm{k}+\bm{q}} \right)^{-1}, \label{eq:end:d}
\end{align}
where Equations from \Cref{eq:start:A}) to \Cref{eq:end:d} are all matrices in spinor degree of freedom. For $\Pi^{\bot, R/K}_{Q,\bm{q}, 00}(\omega)$, simply $B_{kl,\bm{kq}\sigma}$ is replaced by $C_{kl,\bm{kq}\sigma}$. Using the above expressions, we can compute the magnon distribution function $F_q$, which is further discussed in \cref{subsec:distribution_function}.

\subsection{Distribution function}
\label{subsec:distribution_function}
Here, we derive the magnon distribution function $F_q$ with \Cref{eq:PiR_sep,eq:PiK_sep}. 
As in the main text, we only consider the time-averaged component and we can find $2 \mathrm{i} \text{Im} \left[ \chi^{\bot, R}_{0,\bm{q},00}(\omega) \right]$ as follows:
\begin{align}
\bm{\chi}^{\bot, R}_{\bm{q},00}(\omega) & \approx \left[\frac{1}{2I}-\bm{\Pi}^{\bot, R}_{\bm{q}, 00} (\omega)  \right]^{-1} \notag \\
   &=  \frac{1}{\underbrace{  \text{det}\left[(2I)^{-1}-\bm{\Pi}^{\bot, R}_{\bm{q}, 00} (\omega)  \right] }_{\equiv D^{\bot, R}_{\bm{q},00}(\omega)}}
\begin{pmatrix} 
\frac{1}{2I}-\Pi^{\bot, R}_{0, \bm{Q}+\bm{q}} & \Pi^{\bot, R}_{Q,\bm{q}} \\
\Pi^{\bot, R}_{Q,\bm{q}} &  \frac{1}{2I}-\Pi^{\bot, R}_{0,\bm{q}}\\
\end{pmatrix} \\
2 \mathrm{i} \text{Im} \left[ \chi^{\bot, R}_{0,\bm{q},00}(\omega) \right]&=\frac{2 \mathrm{i} \text{Im} \left[D^{\bot, R,\ast}_{\bm{q},00}(\omega) \left(\frac{1}{2I}-\Pi^{\bot, R}_{0, \bm{Q}+\bm{q}}\right) \right] }{|D^{\bot, R}_{\bm{q},00}(\omega)|^2}  \notag \\
&= \frac{R\left[\Pi^{R}_{\text{mag}} \text{~terms}\right] + dR\left[\text{all~} \delta\Pi^{R} \text{relevant terms}\right] }{|D^{\bot, R}_{\bm{q},00}(\omega)|^2} 
\end{align}
Likewise, we can find $\chi^{\bot, K}_{0,\bm{q},00}(\omega)$ as follows:

\begin{align}
\bm{\chi}^{\bot, K}_{\bm{q},00}(\omega) & \approx \bm{\chi}^{\bot, R}_{\bm{q},00}(\omega)  \bm{\Pi}^{\bot, K}_{\bm{q},00} (\omega)  \bm{\chi}^{\bot, A}_{\bm{q},00}(\omega) \\
\chi^{\bot, K}_{0,\bm{q},00}(\omega) &=\frac{K\left[\Pi^{R/K}_{\text{mag}} \text{~terms}\right] + dK\left[\text{all~} \delta\Pi^{R/K} \text{~relevant terms}\right] }{|D^{\bot, R}_{\bm{q},00}(\omega)|^2} 
\end{align}
Using the Fluctuation-dissipation theorem for the Magnus limit expression:
\begin{align}
\frac{K\left[\Pi^{R/K}_{\text{mag}} \text{~terms}\right]}{R\left[\Pi^{R}_{\text{mag}} \text{~terms}\right]}&=\frac{\chi^{\bot, K}_{\text{mag},0,\bm{q},00}(\omega)}{2 \mathrm{i} \text{Im} \left[ \chi^{\bot, R}_{\text{mag},0,\bm{q},00}(\omega) \right]} =F_{\text{mag},q}(\omega) = \mathrm{coth} \frac{\beta}{2}(\omega), 
\end{align}
we can find the magnon distribution function as follows:
\begin{align}
   F_q  
  &\approx\frac{\mathrm{coth} \frac{\beta}{2}(\omega_q) +  \frac{   d K\left(\omega_q, \Omega, \Gamma, \Tilde{t} J_{0}\left(\frac{E}{\Omega}\right), \Tilde{t} J_{1}\left(\frac{E}{\Omega}\right)\right)}{R\left(\omega_q, \Gamma, \Tilde{t} J_{0}\left(\frac{E}{\Omega}\right)\right) }}{1+\frac{ d R\left(\omega_q, \Omega, \Gamma, \Tilde{t} J_{0}\left(\frac{E}{\Omega}\right), \Tilde{t} J_{1}\left(\frac{E}{\Omega}\right)\right)}{R\left(\omega_q, \Gamma, \Tilde{t} J_{0}\left(\frac{E}{\Omega}\right)\right) }}, \label{eq:2ndF_q}
\end{align}
where we explicitly show all dependent valuables for $dK, dR,$ and $R$.
The resultant distribution function is presented in \Cref{fig:expand_ana}. 
In \Cref{fig:full_3panel_Fq}, we also show exact numerical results (the lowest panel is identical to FIG.4 of the main text). Although the divergent feature (interpret as vanishing of $\text{Im}\left[\bm{\chi}_{00,\bm{q},0}\left( \omega_q \right) \right]$ at the non-zero $\omega_q$) is less prominent in \Cref{fig:expand_ana} compared to \Cref{fig:full_3panel_Fq}, the overall feature is well-captured by this approximation.

Because of the even and odd symmetry argument with respect to $\omega=0$, we have in the $\omega \rightarrow 0$ limit:
\begin{align}
   dK &\rightarrow \text{Constant} \\
   dR &\rightarrow \text{Constant} \times \omega \\
   R &\rightarrow \text{Constant} \times \omega  
\end{align}
Therefore, the magnon distribution function exhibits the usual thermal divergence for the $\omega_q \rightarrow 0$ limit:
\begin{align}
    F_q  \rightarrow \frac{\frac{\beta}{2\omega_q} + \frac{C_1}{\omega_q}}{1+C_2} \propto \frac{C}{\omega_q} \label{eq:Fq_wzero_limit}.
\end{align}
where, $C, C_1$, and $C_2$ are all independent of $\omega_q$. We plot the results of $\frac{C}{\omega_q}$ in the inset of the lower panel of \Cref{fig:expand_ana} as line curves, which shows good agreement.

In the zero temperature limit, $\mathrm{coth} \frac{\beta}{2}(\omega_q) \approx 1$ except for $\omega_q=0$. 
Thus, for $dK, dR>>1$, $F_q$ tends to 
\begin{align}
    F_q& \rightarrow \frac{  d K\left(\omega_q, \Omega, \Gamma, \Tilde{t} J_{0}\left(\frac{E}{\Omega}\right), \Tilde{t} J_{1}\left(\frac{E}{\Omega}\right)\right)}{d R\left(\omega_q, \Omega, \Gamma, \Tilde{t} J_{0}\left(\frac{E}{\Omega}\right), \Tilde{t} J_{1}\left(\frac{E}{\Omega}\right)\right)}. \label{eq:dKdR}
\end{align}
The above expression is almost independent of $E$ since $dR, dK \propto |J_1(E/\Omega)|^2$ except the renormalization of the hopping parameter $\Tilde{t} \rightarrow \Tilde{t} J_0(E/\Omega)$ (Magnus limit). This can be seen for the intermediate to high $\omega_q$-regime ($\omega_q >0.2$) in \Cref{fig:expand_ana}. $F_q$ from \Cref{eq:dKdR} and corresponding effective temperature $T_{\text{eff}}$ are plotted in the middle and lower panels of \Cref{fig:expand_ana} as cross markers.

\subsubsection{$\Tilde{t}/\Omega$ expansion}
\label{subsubsec:m1omega_expansion}
We can further approximate the expression by expanding $d\Pi^{R/K}$ with respect to $\Tilde{t}/\Omega$ as we are a high $\Omega$ regime. As an example, we only show the retarded component $d\Pi^{R}$ and focus on the imaginary part since the real part is mainly dictated by $\Pi^{R/K}_{\text{mag}}$. 
We evaluate the following integral at the zero temperature limit:
\begin{align}
    \Tilde{M}^{R}_{jikl,\bm{k q}}(\omega)\equiv\int d \omega' M^{R}_{jikl,\bm{k q}}(\omega^{\prime}, \omega) \label{eq:MR_int}.
\end{align}
Then, we expand the resultant expression with respect to $\Tilde{t}/\Omega$ and the most dominant contributions come from
\begin{align}
  \Tilde{M}^{R}_{jikl,\bm{k q}}(\omega)= -l\frac{8 \Gamma \pi \delta_{jk}\delta_{kl}}{ \left(l \omega-E_{\bm{k}}+E_{\bm{kq}}\right)^2+4 \Gamma^2} \frac{1}{\Tilde{t}^2} \left(\frac{\Tilde{t}}{\Omega}\right)^2 + O\left(\left(\frac{\Tilde{t}}{\Omega}\right)^3  \right) \label{eq:MR_omega2nd}.
\end{align}
Note that the terms with an odd power of $\left(\frac{\Tilde{t}}{\Omega}\right)$ cancel out due to the symmetry of $d_1$ and $d_3$ in the first three terms in $M^{R}_{jikl,\bm{k q}}(\omega^{\prime}, \omega)$. Furthermore, the last two terms in $M^{R}_{jikl,\bm{k q}}(\omega^{\prime}, \omega)$ ($R_{-1}$ and $R_{1}$ terms) have $\left(\frac{\Tilde{t}}{\Omega}\right)^3$-order as their leading order. Since, to order $\left(\frac{\Tilde{t}}{\Omega}\right)^2$-order, $\Tilde{M}^{R}_{jikl,\bm{k q}}(\omega)$ does not depend on index $i$, the sum in \Cref{eq:R_tensor_sum} can be simplified as $\sum_{i}A_{ji, \bm{k}\sigma}A_{ik, \bm{k}\sigma}=\delta_{kj}$. After taking the $k,j$-sum in \Cref{eq:R_tensor_sum}, one obtains 
\begin{align}
    d\Pi^{R}_{\bm{k},\bm{q},\downarrow}(\omega)= d\Pi^{R,\ast}_{\bm{k}+\bm{q},-\bm{q},\uparrow}(-\omega)&=-\frac{16 \pi \omega \Gamma (E_{\bm{k}}-E_{\bm{kq}}) \left(\epsilon_{\bm{k}} \epsilon_{\bm{kq}}+E_{\bm{k}} E_{\bm{kq}}-\left(m_0^{\left(0\right)}\right)^2\right)}{E_{\bm{k}} E_{\bm{kq}} \Omega^2 \left((\omega+E_{\bm{k}}-E_{\bm{kq}})^2+4 \Gamma^2\right) \left((\omega-E_{\bm{k}}+E_{\bm{kq}})^2+4 \Gamma^2\right)}   \\
 \text{Im} \biggl[ \delta\Pi^{\bot, R}_{0,\bm{q}} (\omega)\biggr]&=-\frac{1}{N } \sum^{\prime}_{\bm{k}} \frac{4 \omega \Gamma \left(|\epsilon_{\bm{k},1}|^2-|\epsilon_{\bm{k}+\bm{q},1}|^2\right)(E_{\bm{k}}-E_{\bm{kq}}) \left(\epsilon_{\bm{k}} \epsilon_{\bm{kq}}+E_{\bm{k}} E_{\bm{kq}}-\left(m_0^{\left(0\right)}\right)^2\right)}{E_{\bm{k}} E_{\bm{kq}} \Omega^2 \left((\omega+E_{\bm{k}}-E_{\bm{kq}})^2+4 \Gamma^2\right) \left((\omega-E_{\bm{k}}+E_{\bm{kq}})^2+4 \Gamma^2\right)}  \label{eq:ImdPiR_0q}   
\end{align}
 
For the $Q$ component (off-diagonal), we just need to change $B_{kl, \bm{k q}\sigma}$ to $C_{kl, \bm{k q}\sigma}$. Using \Cref{eq:MR_omega2nd} and carrying out the sum over $j,k$ in \Cref{eq:R_tensor_sum} yields 
\begin{align}
    d\Pi^{R}_{\bm{k},\bm{q},\downarrow}(\omega)= d\Pi^{R,\ast}_{\bm{k}+\bm{q},-\bm{q},\uparrow}(-\omega)&=  \frac{1}{N } \sum^{\prime}_{\bm{k}} \frac{8 \pi \Gamma m_0^{\left(0\right)} (E_{\bm{k}}-E_{\bm{kq}}) \left(\omega^2+(E_{\bm{k}}-E_{\bm{kq}})^2+4 \Gamma^2\right)}{E_{\bm{k}} E_{\bm{kq}} \Omega^2 \left((\omega+E_{\bm{k}}-E_{\bm{kq}})^2+4 \Gamma^2\right) \left((\omega-E_{\bm{k}}+E_{\bm{kq}})^2+4 \Gamma^2\right)}\\
   \text{Im} \biggl[ \delta\Pi^{\bot, R}_{Q,\bm{q}} (\omega)\biggr]&=\frac{1}{N } \sum^{\prime}_{\bm{k}} \frac{2 \Gamma m_0^{\left(0\right)} \left(|\epsilon_{\bm{k},1}|^2-|\epsilon_{\bm{k}+\bm{q},1}|^2\right)(E_{\bm{k}}-E_{\bm{kq}}) \left(\omega^2+(E_{\bm{k}}-E_{\bm{kq}})^2+4 \Gamma^2\right)}{E_{\bm{k}} E_{\bm{kq}} \Omega^2 \left((\omega+E_{\bm{k}}-E_{\bm{kq}})^2+4 \Gamma^2\right) \left((\omega-E_{\bm{k}}+E_{\bm{kq}})^2+4 \Gamma^2\right)}   \label{eq:ImdPiR_Qq}
\end{align}
Note that \Cref{eq:ImdPiR_0q} is always positive while \Cref{eq:ImdPiR_Qq} is negative, which can be understood by applying the large gap $m_0^{\left(0\right)}$ and small $\delta q$ (around $\bm{q} \rightarrow \bm{Q}$) expansion for $\left(|\epsilon_{\bm{k},1}|^2-|\epsilon_{\bm{k}+\bm{q},1}|^2\right)(E_{\bm{k}}-E_{\bm{kq}})$ for both expressions (see also the discussion in \Cref{sec:effective_damping}).
Using \Cref{eq:ImdPiR_0q,eq:ImdPiR_Qq} together with the corresponding Keldysh part, we compute the distribution function $F_q$, which is summarized in \Cref{fig:expandw2_ana}. Although the non-zero $\omega_q$ divergent feature is less prominent compared with \Cref{fig:expand_ana} (threshold $E$ value is smaller in \Cref{fig:expandw2_ana}, e.g $E=10$), qualitative feature is still captured within this approximation. Note that since $T=0$ is used for \Cref{fig:expandw2_ana} whereas $T=0.01$ is employed for \Cref{fig:expand_ana}, the discrepancy observed in the inset of $T_{\text{eff}}$ is also rooted in the $\coth{\left(\frac{\beta\omega_q}{2}\right)}$ contribution in \Cref{eq:2ndF_q}.

\subsubsection{Effective damping}
\label{sec:effective_damping}
\begin{figure}[t!]
\begin{center} 
\includegraphics[width=0.45\textwidth, angle=-0]{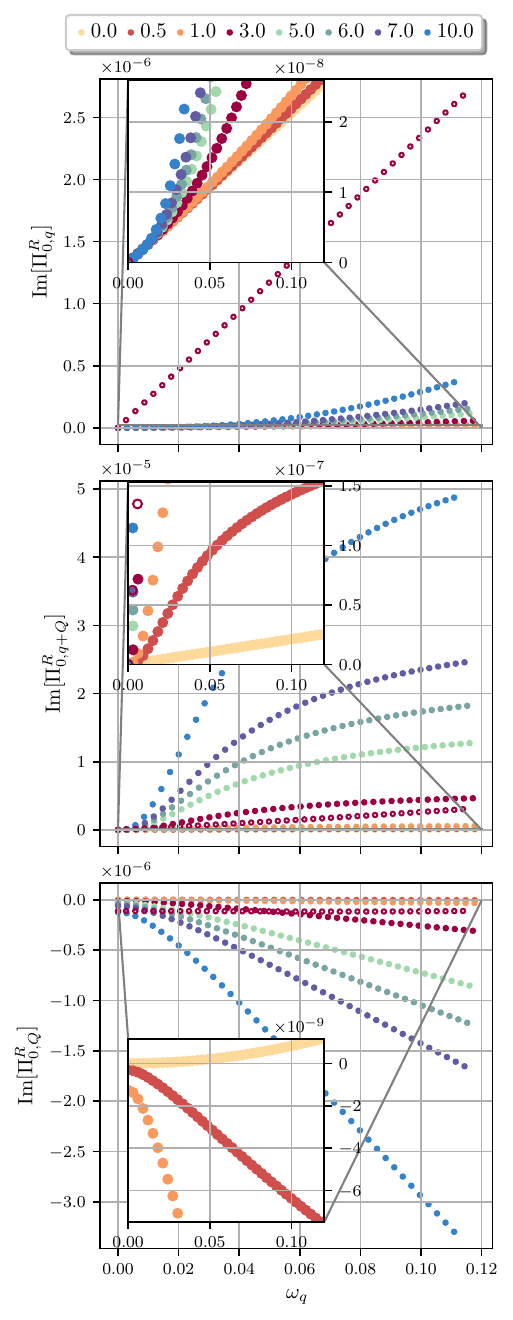}
\caption{Upper, middle, and lower panels show the imaginary part of the time-averaged dressed polarization bubbles, $\text{Im}\left[\Pi^{R}_{0,q, 00} \right]$, $\text{Im}\left[\Pi^{R}_{0,q+Q, 00} \right]$, and $\text{Im}\left[\Pi^{R}_{0,Q, 00} \right]$ as a function of magnon frequency $\omega_q$ for several drive amplitudes $E$, which are indicated in the legend on top of the upper panel. Each inset shows a zoom-in close to zero value of the $y-$axis. The choice of parameters and meaning of the empty circles are the same as in FIG. 1 of the main text.}
\label{fig:4pi}
\end{center}
\end{figure}

Here, we want to demonstrate how and where the sign change occurs in $\text{Im}\left[\left[\chi^{\bot, R}_{0,\bm{q},00}(\omega_q)\right]^{-1}\right]$ found in Section II and III of the main text by employing the analysis introduced in \Cref{sec:Expansion_polarization}. Using the expression in \Cref{eq:PiR_sep,eq:PiK_sep}, we can perform the following separation for $\left[\chi^{\bot, R}_{0,\bm{q},00}(\omega_q)\right]^{-1}$: 
\begin{align}
\bm{\Pi}^{\bot, R}_{\bm{q},00}&=
\begin{pmatrix}
\Pi^{\bot, R}_{0,\bm{q}} & \Pi^{\bot, R}_{Q,\bm{q}} \\
\Pi^{\bot, R}_{Q,\bm{q}} & \Pi^{\bot, R}_{0, \bm{Q}+\bm{q}} \\
\end{pmatrix} =
\bm{\Pi}^{\bot, R}_{\text{mag},\bm{q}}+\bm{\delta \Pi}^{\bot, R}_{\bm{q}} \label{eq:bubble_matrix}\\
\bm{\delta \Pi}^{\bot, R}_{\bm{q}}&=     
\begin{pmatrix}
\delta\Pi^{\bot, R}_{0,\bm{q}} & \delta\Pi^{\bot, R}_{Q,\bm{q}} \\
\delta\Pi^{\bot, R}_{Q,\bm{q}} & \delta\Pi^{\bot, R}_{0, \bm{Q}+\bm{q}} \\
\end{pmatrix}   \\  
\\
  \chi^{\bot, R}_{0,\bm{q},00}(\omega) & = \left[\frac{1}{2I}-\bm{\Pi}^{\bot, R}_{\bm{q},00} (\omega)  \right]^{-1}_{0} \\
  &\equiv \frac{1}{ \left(\chi^{\bot, R}_{\text{mag},0,\bm{q}}(\omega)\right)^{-1} -\Sigma^{\bot, R}_{\bm{q}}(\omega)    }  \\
  \left(\chi^{\bot, R}_{\text{mag},0,\bm{q}}(\omega)\right)^{-1}&=\left(-\Tilde{\Pi}_{\text{mag},0,\bm{q}}(\omega)\right)-  \frac{\left(-\Tilde{\Pi}^{\bot, R}_{\text{mag},Q,\bm{q}}(\omega)\right)^2}{\left(-\Tilde{\Pi}^{\bot, R}_{\text{mag},0,\bm{Q}+\bm{q}}(\omega)\right)}  \\
  \Sigma^{\bot, R}_{\bm{q}}(\omega)
  &\approx -\left(-\delta \Pi^{\bot, R}_{0,\bm{q}}(\omega) \right) -   \frac{\left(-\Tilde{\Pi}^{\bot, R}_{\text{mag},Q,\bm{q}}(\omega)\right)^{2}}{\left(-\Tilde{\Pi}^{\bot, R}_{\text{mag},0,\bm{Q}+\bm{q}}(\omega)\right)^{2}}\left(-\delta \Pi^{\bot, R}_{0,\bm{Q}+\bm{q}}(\omega)\right)   \notag \\
  &+ 2 \frac{\left(-\Tilde{\Pi}^{\bot, R}_{\text{mag},Q,\bm{q}}(\omega)\right)}{ \left(-\Tilde{\Pi}^{\bot, R}_{\text{mag},0,\bm{Q}+\bm{q}}(\omega)\right)}\left(-\delta \Pi^{\bot, R}_{Q,\bm{q}}(\omega) \right)+O\left( \left(J_1\left(\frac{E}{\Omega}\right)\right)^3 \right) \label{eq:Self_2nd}
\end{align}  
where we introduced the following notations:
\begin{align}
\Tilde{\Pi}^{\bot, R}_{\text{mag},0,\bm{q}}(\omega)&\equiv \Pi^{\bot, R}_{\text{mag},0,\bm{q}} (\omega)  - \frac{1}{2 I}, \\
\Tilde{\Pi}^{\bot, R}_{\text{mag},Q,\bm{q}}(\omega)&\equiv \Pi^{\bot, R}_{\text{mag},Q,\bm{q}} (\omega).  
\end{align}

As in Section II of the main text, the effective damping can be found as follows:
\begin{align}
\Tilde{\Gamma}_q\equiv\frac{\Gamma_q^{\text{eff}}}{Z_q}&=-\text{Im}\left[\left[\chi^{\bot, R}_{\bm{q},00,0}(\omega_q)\right]^{-1}\right] \\
&=-\text{Im}\left[\left(\chi^{\bot, R}_{\text{mag},0,\bm{q}}\right)^{-1} - \Sigma^{\bot, R}_{\bm{q}}\right] \label{eq:Gamma_til} \\
-\text{Im}\left[\left(\chi^{\bot, R}_{\text{mag},0,\bm{q}}\right)^{-1}\right]&=\underbrace{\text{Im}\left[\Tilde{\Pi}_{\text{mag},0,\bm{q}}\right]}_{\Tilde{\Gamma}_{q,1}^{\text{eq}}}-\text{Im}\left[\left(\Tilde{\Pi}^{\bot, R}_{\text{mag},Q,\bm{q}}\right)^2 \left(\Tilde{\Pi}^{\bot, R}_{\text{mag},0,\bm{Q}+\bm{q}}\right)^{-1} \right] \label{eq:Imchimag}\\
-\text{Im}\left[\left(\Tilde{\Pi}^{\bot, R}_{\text{mag},Q,\bm{q}}\right)^2 \left(\Tilde{\Pi}^{\bot, R}_{\text{mag},0,\bm{Q}+\bm{q}}\right)^{-1} \right]&=-\text{Re}\left[\Tilde{\Pi}^{\bot, R}_{\text{mag},Q,\bm{q}}\right]^2 \text{Im}\left[\left(\Tilde{\Pi}^{\bot, R}_{\text{mag},0,\bm{Q}+\bm{q}}\right)^{-1} \right] \\
&-2\text{Re}\left[\Tilde{\Pi}^{\bot, R}_{\text{mag},Q,\bm{q}}\right] \text{Re}\left[\left(\Tilde{\Pi}^{\bot, R}_{\text{mag},0,\bm{Q}+\bm{q}}\right)^{-1} \right]\text{Im}\left[\Tilde{\Pi}^{\bot, R}_{\text{mag},Q,\bm{q}}\right]\\
&\equiv \Tilde{\Gamma}_{q,2}^{\text{eq}}+\Tilde{\Gamma}_{q,3}^{\text{eq}} \\
\text{Im}\left[\Sigma^{\bot, R}_{\bm{q}}\right]&\approx\text{Im}\left[\delta \Pi^{\bot, R}_{0,\bm{q}} \right] +   \text{Re}\left[\left(\Tilde{\Pi}^{\bot, R}_{\text{mag},Q,\bm{q}}\right)^{2}\right] \text{Re}\left[\left(\Tilde{\Pi}^{\bot, R}_{\text{mag},0,\bm{Q}+\bm{q}}\right)^{-2}\right] \text{Im}\left[\delta \Pi^{\bot, R}_{0,\bm{Q}+\bm{q}}\right] \notag \\
  & - 2 \text{Re}\left[\Tilde{\Pi}^{\bot, R}_{\text{mag},Q,\bm{q}}\right] \text{Re}\left[\left(\Tilde{\Pi}^{\bot, R}_{\text{mag},0,\bm{Q}+\bm{q}}\right)^{-1}\right]\text{Im}\left[\delta \Pi^{\bot, R}_{Q,\bm{q}} \right] \label{eq:Imself_2nd}\\
  & \equiv  \Tilde{\Gamma}_{q,1}^{\text{neq}}+\Tilde{\Gamma}_{q,2}^{\text{neq}}+\Tilde{\Gamma}_{q,3}^{\text{neq}} 
\end{align}
where we only kept the dominant real and imaginary combinations for each term. 
By numerically analyzing each contribution, we found
\begin{align*}
    \Tilde{\Gamma}_{q,1}^{\text{eq}},\Tilde{\Gamma}_{q,2}^{\text{eq}},\Tilde{\Gamma}_{q,3}^{\text{eq}},\Tilde{\Gamma}_{q,1}^{\text{neq}},\Tilde{\Gamma}_{q,2}^{\text{neq}}&>0\\
    \Tilde{\Gamma}_{q,3}^{\text{neq}}&<0
\end{align*}
In particular, the most curious point is $\Tilde{\Gamma}_{q,3}^{\text{eq}}>0$ while $\Tilde{\Gamma}_{q,3}^{\text{neq}}<0$, which originated from $\text{Im}\left[\Tilde{\Pi}^{\bot, R}_{\text{mag},Q,\bm{q}}\right]>0$ while $\text{Im}\left[\delta \Pi^{\bot, R}_{Q,\bm{q}} \right]<0$ as we briefly mentioned in \cref{subsubsec:m1omega_expansion}. 
If $\text{Im}\left[\delta \Pi^{\bot, R}_{Q,\bm{q}} \right]>0$, the sign change would not occur in $\Tilde{\Gamma}$. In \Cref{fig:4pi}, we show each component of the imaginary part of the polarization bubble in \Cref{eq:bubble_matrix}, which clearly demonstrates that the driving simply increases the dissipation for the diagonal component $\text{Im}\left[\Pi^{\bot, R}_{0,\bm{q}} \right]$ and $\text{Im}\left[\Pi^{\bot, R}_{0,\bm{q}+\bm{Q}} \right]$ while the off diagonal component $\text{Im}\left[\Pi^{\bot, R}_{Q,\bm{q}} \right]$ is decreased. As soon as the driving is on, the sign of $\text{Im}\left[\Pi^{\bot, R}_{Q,\bm{q}} \right]$ changes (see the inset of the lower panel).

\begin{figure}[t!]
\begin{center} 
\includegraphics[width=0.95\textwidth, angle=-0]{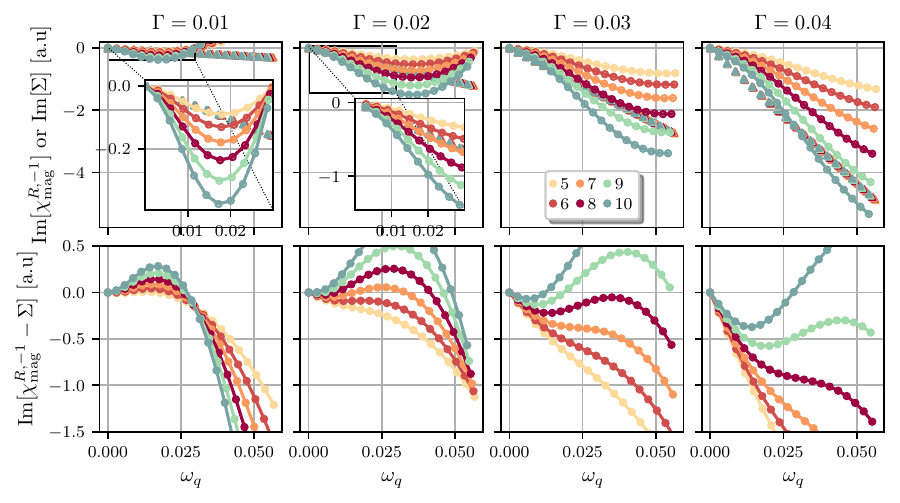}
\caption{Upper panels show $\text{Im}\left[\left(\chi^{\bot, R}_{\text{mag},0,\bm{q}}\right)^{-1}\right]$ (triangles) and $\text{Im}\left[\Sigma\right]$ (connected empty circles) for several $E$ and $\Gamma$ from \Cref{eq:Gamma_til}. The parameter sets $E$ and $\Gamma$ are indicated in the legend of the top third panel and on top of each panel, respectively. Here, $\Sigma$ is obtained as $\Sigma=\chi^{R,-1}_{\text{mag}}-\chi^{R,-1}$. The lower panels show the subtraction of the above panels with positive sign signifying the instability exactly as in FIG. 3 of the main text. The parameter sets are the same as in FIG. 4 of the main text.}
\label{fig:effGamma}
\end{center}
\end{figure}

Additionally, we show the separated contributions for $\Tilde{\Gamma}$ as in \Cref{eq:Gamma_til} in \Cref{fig:effGamma} where the lower panels are exactly the same as in FIG.4 of the main text. The Magnus limit term $\chi^{R,-1}_{\text{mag}}$ is almost independent of the drive amplitude $E$ and proportional to $\propto -\Gamma^2 \omega_q$ in the small $\omega_q$ regime (plotted $\omega_q$ window). $\text{Im}\left[\Sigma\right]$ drives the system toward the instability due to the sign of $\text{Im}\left[\Sigma\right]$. $\text{Im}\left[\Sigma\right]$ has minimum value at non-zero $\omega_q$, which is determined by $\Gamma$. Also, $\text{Im}\left[\Sigma\right]$ is almost proportional to $\propto (E/\Omega)^2$ as in \Cref{eq:Self_2nd}. Therefore, the instability only occurs for large $E/\Omega$ and small $\Gamma/\Tilde{t}$.     

\subsection{Origin of the Instability}
\label{sec:origin_of_instability_SI}
In this section, we show how $\bm{\delta \Pi}_{\bm{q}}^{\bot,R}$ contributes to $\bm{\Pi}_{\bm{q},00}^{\bot,R}$ in \Cref{eq:bubble_matrix}. In particular, we demonstrate how $\text{Im}\left[\bm{\delta \Pi}_{\bm{q}}^{\bot,R}\right]$ is unique compared to $\text{Im}\left[\bm{\Pi}_{\bm{q},\text{mag}}^{R}\right]$ (Magnus limit contribution) as explained in the Section IV of the main text. We present the root cause why $\text{Im}\left[\Tilde{\Pi}^{\bot, R}_{\text{mag},Q,\bm{q}}\right]>0$ whereas $\text{Im}\left[\delta \Pi^{\bot, R}_{Q,\bm{q}} \right]<0$, as discussed in \Cref{sec:effective_damping}.    
\subsubsection{Time-averaged Green's function}
Here we use the same argument as in \Cref{sec:Expansion_polarization}, but without expanding with respect to $\epsilon_{\bm{k},1}/\Tilde{t}$. Then, the time-averaged retarded Green's function in \Cref{eq:GR_tridigonal} is given as  
\begin{align}
G_{00}^{R}&=\left(M_{00}-M_{10}^{\dagger} M_{11}^{-1} M_{10}-M_{0-1} M_{-1-1}^{-1} M_{0-1}^{\dagger}\right)^{-1}.     
\end{align}  
For $\Omega\gg E_{\bm{k}}$ ($\omega\in[-\infty,\infty]$ is the integration variable), we can approximate terms originating from the Floquet sectors $\pm 1$ ($M_{11}$ and $M_{-1-1}$) as follows:  
\begin{align}
   d_{n = \pm1, \bm{k}}^{\pm}&=\left(\omega +n \Omega + \mathrm{i} \Gamma \mp E_{\bm{k}} \right)^{-1} \approx \left(\omega +n \Omega + \mathrm{i} \Gamma \right)^{-1} \\
 M_{10}^{\dagger} M_{11}^{-1} M_{10}+M_{0-1} M_{-1-1}^{-1} M_{0-1}^{\dagger} &=  |\epsilon_{\bm{k},1}|^2  \sum_{n=-1,1} \tau_3 \left(U_{\bm{k},\sigma} U_{\bm{k},\sigma}^{\dagger} M_{nn,\bm{k}\sigma} U_{\bm{k},\sigma} U_{\bm{k},\sigma}^{\dagger} \right)^{-1} \tau_3 \notag\\
 &=|\epsilon_{\bm{k},1}|^2 \sum_{n=-1,1} \tau_3 \left(U_{\bm{k},\sigma} d_{n,\bm{k}}^{\pm}  U_{\bm{k},\sigma}^{\dagger} \right)\tau_3 \notag\\
 &\approx |\epsilon_{\bm{k},1}|^2 \sum_{n=-1,1} \left(\omega +n \Omega + \mathrm{i} \Gamma \right)^{-1} \\
 &= \sum_{n=-1,1} \Sigma_{n,\bm{k}}(\omega) \notag\\
 \Sigma_{n,\bm{k}}(\omega) &\equiv \frac{|\epsilon_{\bm{k},1}|^2}{\omega +n \Omega + \mathrm{i} \Gamma } 
\end{align}
Using the above $G^{R}_{00}$, one obtains
\begin{align}
\Pi^{+ -, R}_{0/Q,\bm{q}, 0}\left(\omega\right)&=-\frac{1}{4 \pi \mathrm{i} N} \sum_{\bm{k}} \int d\omega' \mathrm{Tr}\left[G^{R}_{00,\bm{k},\downarrow}\left(\omega^{\prime}\right) (\tau_{0/1})G_{00, \bm{k}+\bm{q},\uparrow}^{K}(\omega^{\prime}-\omega) \right. \notag \\
&\left. + G_{00, \bm{k}, \downarrow}^{K}\left(\omega^{\prime}\right) ( \tau_{0/1})   G_{00, \bm{k}+\bm{q}, \uparrow}^{A}(\omega^{\prime}-\omega)   \right] \\
G_{00, \bm{k}, \downarrow}^{K}\left(\omega^{\prime}\right) &=\underbrace{-2 i \Gamma \tanh{\frac{\beta \omega'}{2}}}_{R_{0,\bm{k}}} G_{00, \bm{k}, \downarrow}^{R}\left(\omega^{\prime}\right) G_{00, \bm{k}, \downarrow}^{A}\left(\omega^{\prime}\right) \label{eq:GK00_GR00GA00}\\
\Pi^{+ -, R}_{0,q, 00}\left(\omega\right)&=-\frac{1}{4 \pi \mathrm{i} N} \sum_{\bm{k},kl} \int d\omega' B_{kq,kl,\downarrow} B_{\bm{kq},lk,\downarrow}^{\dagger} \left(R_{0,\bm{k}+\bm{q}}d^{k}_{2,\bm{k}} d^{l}_{2,\bm{k}+\bm{q}} d^{l,\ast}_{2,\bm{k}+\bm{q}} +R_{0,\bm{k}} d^{k}_{2,\bm{k}} d^{k,\ast}_{2,\bm{k}} d^{l,\ast}_{2,\bm{k}+\bm{q}}\right) \\
\Pi^{+ -, R}_{Q,q, 00}\left(\omega\right)&=-\frac{1}{4 \pi \mathrm{i} N} \sum_{\bm{k},kl} \int d\omega' C_{\bm{kq},kl\downarrow} B_{\bm{kq},lk,\downarrow}^{\dagger} \underbrace{\left(R_{0,\bm{k}+\bm{q}}d^{k}_{2,\bm{k}} d^{l}_{2,\bm{k}+\bm{q}} d^{l,\ast}_{2,\bm{k}+\bm{q}} +R_{0,\bm{k}} d^{k}_{2,\bm{k}} d^{k,\ast}_{2,\bm{k}} d^{l,\ast}_{2,\bm{k}+\bm{q}}\right) }_{M_{\bm{kq}}^{kl}}\\
d^{k}_{2,\bm{k}}&=\frac{1}{\omega'-E_{\bm{k}}^{k}+\mathrm{i}\Gamma-\Sigma_{-1,1,\bm{k}}(\omega')}\\
d^{l}_{2,\bm{k}+\bm{q}}&=\frac{1}{\omega'-\omega-E_{\bm{k}+\bm{q}}^{l}+\mathrm{i}\Gamma-\Sigma_{-1,1,\bm{k}+\bm{q}}(\omega'-\omega)}\\
\Sigma_{-1,1,\bm{k}}(\omega)&\equiv \sum_{n=\pm1}\frac{|\epsilon_{\bm{k},1}|^2}{\omega+n\Omega+i\Gamma}
\end{align}
written in terms of the matrices $C_{\bm{kq},kl\downarrow}$ and $B_{\bm{kq},lk,\downarrow}$ introduced in \Cref{sec:Expansion_polarization}.
Since $C_{\bm{kq},kl\downarrow}$ and $B_{\bm{kq},lk,\downarrow}$ are real, $\text{Im} \left[\Pi^{+ -, R}_{0,q/Q, 00}\left(\omega\right)\right]$ depends only on the real part, $\text{Re}[M_{\bm{kq}}^{kl}]$. Since $R_{0,\bm{k}}$ is imaginary, one should consider the following terms:   
\begin{align}
 \text{Im}[d^{k}_{2,\bm{k}} d^{l}_{2,\bm{k}+\bm{q}} d^{l,\ast}_{2,\bm{k}+\bm{q}}]&=\frac{-\Gamma_{\bm{k}}(\omega')}{\left(\omega'-E_{\bm{k}}^{k}(\omega')\right)^2+\Gamma_{\bm{k}}(\omega')^2}\frac{1}{{\left(\omega'-\omega-E_{\bm{k}+\bm{q}}^{l}(\omega'-\omega)\right)^2+\Gamma_{\bm{k}+\bm{q}}(\omega'-\omega)^2}} \label{eq:first_drive_reno}\\
 \text{Im}[d^{k}_{2,\bm{k}} d^{k,\ast}_{2,\bm{k}} d^{l,\ast}_{2,\bm{k}+\bm{q}}]&=\frac{1}{\left(\omega'-E_{\bm{k}}^{k}(\omega')\right)^2+\Gamma_{\bm{k}}(\omega')^2}\frac{\Gamma_{\bm{k}+\bm{q}}(\omega'-\omega)}{{\left(\omega'-\omega-E_{\bm{k}+\bm{q}}^{l}(\omega'-\omega)\right)^2+\Gamma_{\bm{k}+\bm{q}}(\omega'-\omega)^2}}, \label{eq:second_drive_reno}
\end{align}
where we defined 
\begin{align}
\Gamma_{\bm{k}}(\omega)&\equiv\Gamma-\text{Im}\left[\Sigma_{-1,1,\bm{k}}(\omega) \right] = \Gamma \left(1 + 
\sum_{n=\pm1}\frac{|\epsilon_{\bm{k},1}|^2}{(\omega+n\Omega)^2+\Gamma^2}\right)>\Gamma\\
E_{\bm{k}}^{k}(\omega)&\equiv E_{\bm{k}}^{k}+\text{Re}\left[\Sigma_{-1,1,\bm{k}}(\omega) \right] = E_{\bm{k}}^{k}+\sum_{n=\pm1}\frac{|\epsilon_{\bm{k},1}|^2(\omega+n\Omega)}{(\omega+n\Omega)^2+\Gamma^2}
\end{align}
From this, one arrives at 
\begin{align}
    -\text{Re}[M_{\bm{kq}}^{kl}]&=2\Gamma\frac{\Gamma_{\bm{k}}(\omega')\tanh{\frac{\beta(\omega'-\omega)}{2}}-\Gamma_{\bm{k}+\bm{q}}(\omega'-\omega)\tanh{\frac{\beta(\omega')}{2}}}{\left[{\left(\omega'-\omega-E_{\bm{k}+\bm{q}}^{l}(\omega'-\omega)\right)^2+\Gamma_{\bm{k}+\bm{q}}(\omega'-\omega)^2}\right]\left[\left(\omega'-E_{\bm{k}}^{k}(\omega')\right)^2+\Gamma_{\bm{k}}(\omega')^2\right]} \\
    &\xrightarrow{T\rightarrow0}  2\Gamma\frac{\Gamma_{\bm{k}}(\omega')\sgn{(\omega'-\omega)}-\Gamma_{\bm{k}+\bm{q}}(\omega'-\omega)\sgn{(\omega')}}{\left[{\left(\omega'-\omega-E_{\bm{k}+\bm{q}}^{l}(\omega'-\omega)\right)^2+\Gamma_{\bm{k}+\bm{q}}(\omega'-\omega)^2}\right]\left[\left(\omega'-E_{\bm{k}}^{k}(\omega')\right)^2+\Gamma_{\bm{k}}(\omega')^2\right]} \label{eq:ReMkq}
\end{align}

\subsubsection{Without driving}
In equilibrium, \Cref{eq:ReMkq} yields in the zero-temperature limit 
\begin{align}
   -\text{Re}[M_{\bm{kq}}^{kl}] \xrightarrow{T\rightarrow0}    &= 2\Gamma^2\frac{\sgn(\omega'-\omega)-\sgn(\omega')}{\left[{\left(\omega'-\omega-E_{\bm{k}+\bm{q}}^{l}\right)^2+\Gamma^2}\right]\left[\left(\omega'-E_{\bm{k}}^{k}\right)^2+\Gamma^2\right]} \label{eq:ReM_equiv} \\
  &= \begin{cases} 
      -4\Gamma^2\frac{1}{\left[{\left(\omega'-\omega-E_{\bm{k}+\bm{q}}^{l}\right)^2+\Gamma^2}\right]\left[\left(\omega'-E_{\bm{k}}^{k}\right)^2+\Gamma^2\right]}<0  & 0<\omega'<\omega \\
      0 & \omega'<0, \omega'>\omega 
   \end{cases} 
\end{align}

\subsubsection{With driving}
For $0<\omega'<\omega$ in the presence of driving, the zero-temperature limit becomes
\begin{align}
   -\text{Re}[M_{\bm{kq}}^{kl}] \xrightarrow{T\rightarrow0}    &=\Gamma\frac{\overbrace{\left(\sgn(\omega'-\omega)-\sgn(\omega')\right)}^{-2}\left[\Gamma_{\bm{k}}(\omega')+\Gamma_{\bm{k}+\bm{q}}(\omega'-\omega)\right] }{\left[{\left(\omega'-\omega-E_{\bm{k}+\bm{q}}^{l}(\omega'-\omega)\right)^2+\Gamma_{\bm{k}+\bm{q}}(\omega'-\omega)^2}\right]\left[\left(\omega'-E_{\bm{k}}^{k}(\omega')\right)^2+\Gamma_{\bm{k}}(\omega')^2\right]}<0 \\
   &\approx \Gamma\frac{\overbrace{\left(\sgn(\omega'-\omega)-\sgn(\omega')\right)}^{-2}\left[\Gamma_{\bm{k}}+\Gamma_{\bm{k}+\bm{q}}\right] }{\left[{\left(\omega'-\omega-E_{\bm{k}+\bm{q}}^{l}\right)^2+\Gamma_{\bm{k}+\bm{q}}^2}\right]\left[\left(\omega'-E_{\bm{k}}^{k}\right)^2+\Gamma_{\bm{k}}^2\right]} \label{eq:ReMwp_close_w}
\end{align}
Note that 
\begin{align}
  \Gamma_{\bm{k}}(\omega)&=\Gamma+\sum_{n=\pm 1}\frac{\Gamma |\epsilon_{\bm{k},1}|^2}{\left(\omega+   n \Omega \right)^2+ \Gamma^2  } \notag \\
  &\approx \Gamma+ \frac{2 \Gamma |\epsilon_{\bm{k},1}|^2}{\Omega^2} \equiv \Gamma_{\bm{k}} \\ 
  E_{\bm{k}}^{k}(\omega)&= E_{\bm{k}}^{k}+\sum_{n=\pm1}\frac{|\epsilon_{\bm{k},1}|^2(\omega+n\Omega)}{(\omega+n\Omega)^2+\Gamma^2} \notag \\
  &\approx E_{\bm{k}}^{k}, 
\end{align}
where we used that $\Omega, E_{\bm{k}}\gg\omega, \omega', \Gamma$, which in turn allows us to ignore the $\omega-$dependence. Furthermore, since $\Gamma_{\bm{k}} \approx \Gamma$, we could write \Cref{eq:ReMwp_close_w} as follows:
\begin{align}
   -\text{Re}[M_{\bm{kq}}^{kl}] &\approx 2\Gamma^2 \left[1 +\frac{|\epsilon_{\bm{k},1}|^2+|\epsilon_{\bm{k}+\bm{q},1}|^2}{\Omega^2} \right] \frac{\overbrace{\left(\sgn(\omega'-\omega)-\sgn(\omega')\right)}^{-2} }{\left[{\left(\omega'-\omega-E_{\bm{k}+\bm{q}}^{l}\right)^2+\Gamma^2}\right]\left[\left(\omega'-E_{\bm{k}}^{k}\right)^2+\Gamma^2\right]} \label{eq:noneq_Meq_0wpw}
\end{align}
This is the usual undriven expression found in \Cref{eq:ReM_equiv} plus driving correction ($1\ll\frac{|\epsilon|^2}{\Omega^2}$).

However, for the other $\omega'$ domain (while this integral vanishes for the case of equilibrium), we also have non-zero contributions as follows: 
\begin{align}
   -\text{Re}[M_{\bm{kq}}^{kl}] \xrightarrow{T\rightarrow0}    &=\frac{\Gamma\overbrace{\left(\sgn(\omega'-\omega)+\sgn(\omega')\right)}^{\pm2}\left[  \Gamma_{\bm{k}}(\omega')-\Gamma_{\bm{k}+\bm{q}}(\omega'-\omega) \right]}{\left[{\left(\omega'-\omega-E_{\bm{k}+\bm{q}}^{l}(\omega'-\omega)\right)^2+\Gamma_{\bm{k}+\bm{q}}(\omega'-\omega)^2}\right]\left[\left(\omega'-E_{\bm{k}}^{k}(\omega')\right)^2+\Gamma_{\bm{k}}(\omega')^2\right]} \notag\\
  &=\frac{\Gamma \left(\sgn(\omega'-\omega)+\sgn(\omega')\right) \text{Im}\left[    -\Sigma_{-1,1,\bm{k}}(\omega')+\Sigma_{-1,1,\bm{k}+\bm{q}}(\omega'-\omega) \right]}{\left[{\left(\omega'-\omega-E_{\bm{k}+\bm{q}}^{l}(\omega'-\omega)\right)^2+\Gamma_{\bm{k}+\bm{q}}(\omega'-\omega)^2}\right]\left[\left(\omega'-E_{\bm{k}}^{k}(\omega')\right)^2+\Gamma_{\bm{k}}(\omega')^2\right]} \notag\\
  &\xrightarrow{|\epsilon_{\bm{k},1}|^2 \text{order}} \frac{\Gamma\left(\sgn(\omega'-\omega)+\sgn(\omega')\right) \text{Im}\left[    -\Sigma_{-1,1,\bm{k}}(\omega')+\Sigma_{-1,1,\bm{k}+\bm{q}}(\omega'-\omega) \right]}{\left[{\left(\omega'-\omega-E_{\bm{k}+\bm{q}}^{l}\right)^2+\Gamma^2}\right]\left[\left(\omega'-E_{\bm{k}}^{k}\right)^2+\Gamma^2\right]}  \notag \\
  &= \Gamma^2 \sum_{n=\pm 1}\left(\frac{ |\epsilon_{\bm{k},1}|^2}{\left(\omega'+   n \Omega \right)^2+ \Gamma^2  } - \frac{ |\epsilon_{\bm{k+q},1}|^2}{\left(\omega'-\omega+   n \Omega \right)^2+ \Gamma^2  } \right)\frac{ \left(\sgn(\omega'-\omega)+\sgn(\omega')\right) }{\left[{\left(\omega'-\omega-E_{\bm{k}+\bm{q}}^{l}\right)^2+\Gamma^2}\right]\left[\left(\omega'-E_{\bm{k}}^{k}\right)^2+\Gamma^2\right]}   \label{eq:sgn(omega'-omega)} \\
  & \equiv -\text{Re}[M_{kq}^{\text{2nd},kl}] \notag
\end{align}
\Cref{eq:sgn(omega'-omega)} is the one introduced as the lower case in Equation (10) of the main text. 
Additionally, the $\omega^{\prime}$-integral of \Cref{eq:sgn(omega'-omega)}, followed by a $\Tilde{t}/\Omega$-expansion yields
\begin{align}
   \int_{-\infty}^{\infty} d \omega' \left(-\text{Re}[M_{kq}^{\text{2nd},kl}]\right) 
   &=\frac{|\epsilon_{\bm{k},1}|^2-|\epsilon_{\bm{k}+\bm{q},1}|^2}{\Omega^2} \frac{ 8 \Gamma \pi }{ \left(l \omega-E_{\bm{k}}+E_{\bm{kq}}\right)^2+4 \Gamma^2}\delta_{kl} l + O\left(\left(\frac{\Tilde{t}}{\Omega}\right)^3  \right) 
\end{align}    
This is exactly the same expression as in \Cref{eq:MR_omega2nd}, as it should. 

As emphasized in the main text, the drive-induced breaking of the fluctuation-dissipation theorem  (FDT) renders the contribution of \Cref{eq:sgn(omega'-omega)} non-vanishing. In this light, \Cref{eq:GK00_GR00GA00} can be written as 
\begin{align}
    G_{00, \bm{k}, \downarrow}^{K}\left(\omega^{\prime}\right) &=2i \Im \left[G_{00, \bm{k}, \downarrow}^{R} \left(\omega^{\prime}\right) \right] \frac{\Gamma}{\Gamma_{\bm{k}} \left( \omega'\right)} \tanh{\left(\frac{\beta \omega'}{2}\right)}  \\
    &\neq 2i \Im \left[G_{00, \bm{k}, \downarrow}^{R} \left(\omega^{\prime}\right) \right] \tanh{\left(\frac{\beta \omega'}{2}\right)}. ~~(\text{FDT}) \label{eq:GK00_FDT}
\end{align}
Importantly, if the FDT were satisfied and one would replace \Cref{eq:GK00_GR00GA00} with \Cref{eq:GK00_FDT}:
\begin{align}
    G_{00, \bm{k}, \downarrow}^{K}\left(\omega^{\prime}\right) &= -\frac{2i \Gamma_{\bm{k}}(\omega') \tanh{\left(\frac{\beta \omega'}{2}\right)}} {\left(\omega'-E_{\bm{k}}^{k}(\omega')\right)^2+\Gamma_{\bm{k}}(\omega')^2},
\end{align}
\Cref{eq:ReMkq} would be
\begin{align}
  -\text{Re}[M_{\bm{kq}}^{kl}]&=2\frac{\Gamma_{\bm{k}}(\omega')\Gamma_{\bm{k}+\bm{q}}(\omega'-\omega) \left(\sgn{(\omega'-\omega)}-\sgn{(\omega')}\right)}{\left[{\left(\omega'-\omega-E_{\bm{k}+\bm{q}}^{l}(\omega'-\omega)\right)^2+\Gamma_{\bm{k}+\bm{q}}(\omega'-\omega)^2}\right]\left[\left(\omega'-E_{\bm{k}}^{k}(\omega')\right)^2+\Gamma_{\bm{k}}(\omega')^2\right]}.   
\end{align}
A non-zero contribution arises only from $0<\omega'<\omega$-domain integral. Within our parameter regime ($\Omega, E_{\bm{k}}\gg\omega, \omega', \Gamma$), one recovers the same expression as \Cref{eq:noneq_Meq_0wpw}, i.e., the usual undriven result with an additional correction due to the Floquet drive.

\subsection{Vanishing of the Effective Damping}
\label{subsec:damping}
In previous sections, we connected the sign change of the effective damping $\Gamma_{q}^{\text{eff}}$ with the onset of instabilities in the system. Here, we briefly present a mathematical derivation of this connection. For simplicity, we express the time-averaged transverse fluctuation propagator \Cref{eq:chiR_original} in terms of the time-averaged polarization bubble \Cref{eq:R_pi_matrix}:
\begin{align}
   \chi^{\bot, R}_{0,\bm{q},00}(\omega) & \approx \left[\frac{1}{2I}-\bm{\Pi}^{\bot, R}_{\bm{q},00} (\omega)  \right]^{-1}_{0} \label{eq:TAchi_TApi}. 
\end{align}
We consider the effective action for the antiferromagnetic fluctuation field $\delta m^{c/q}$~, where $c/q$ denotes the classical and quantum components \cite{Kamenev_2023, altland_simons_2010}. For simplicity, we ignore the Keldysh component in the action. At the quadratic order in $\delta m$, we integrate out the quantum component of $\delta m^{q}$, yielding~\cite{Mitra2006, coleman_2015, altland_simons_2010} (note that 
Keldysh component would produce a noise term in \Cref{eq:after_integ_dm} encoding quantum and thermal fluctuations):
\begin{align}
&\left[\frac{1}{2I}-\bm{\Pi}^{\bot, R}_{\bm{q}, 00} (\omega)  \right] \label{eq:after_integ_dm}
\begin{pmatrix}
  \delta m_{-\bm{q},0}^{+,c}(\omega) \\
  \delta m_{-\bm{q}-\bm{Q},0}^{+,c}(\omega)  
\end{pmatrix} 
=\bm{0},
\end{align}
where $\delta m^{\pm}=\delta m^x\pm\mathrm{i}\delta m^y$ and index $0$ denotes the time-averaged component. Solving for $\delta m_{-\bm{q},0}^{+,c}(\omega)$, we obtain
\begin{align}
\left[\chi^{R,\bot}_{0,\bm{q},00}(\omega)\right]^{-1} \delta m_{-\bm{q},0}^{+,c}(\omega) =0, \label{eq:chiRinv_dm_0} 
\end{align}
where we have used \Cref{eq:TAchi_TApi}. Close to the pole $\omega\approx \omega_{q}-i \Gamma_{q}^{\text{eff}}$, we employ  
\begin{align}
  \left[\chi^{R,\bot}_{0,\bm{q},00}(\omega)\right]^{-1}  \approx& \pi Z_{q}^{-1} \left( \omega_{q}-\omega  -i \Gamma_{q}^{\text{eff}} \right) \label{eq:chi_fit}
\end{align}
as is introduced in Equation (5) of the main text. 
Performing the Fourier transform to time ($\omega \rightarrow i\partial_t$), \Cref{eq:chiRinv_dm_0} becomes 
\begin{align}
    \pi Z_{q}^{-1} \left[    \omega_{q}   -  i\left(\partial_t +\Gamma_{q}^{\text{eff}}\right)    \right]\delta m_{-\bm{q},0}^{+,c}(t)=0. \label{eq:gam_t_dm_0}
\end{align}
The solution for the fluctuation $\delta m$ is given as  
\begin{align}
   \delta m_{-\bm{q},0}^{+,c}(t)  \propto  e^{-\left(\Gamma_{q}^{\text{eff}}+ i \omega_{q}   \right)t}.  
\end{align}
Therefore, a negative $\Gamma_{q}^{\text{eff}}$ corresponds to exponentially growing fluctuation in time. In the present model, the sign change occurs for non-zero momentum $\bm{q}$ and frequency $\omega_{q}$. Therefore, this instability falls under the IIIs scenario of Cross and Hohenberg~\cite{Cross1993}, which can potentially lead to time-periodic spatial pattern formation. 
The final state toward which this instability evolves remains an open question for our future work. Addressing this requires to consider higher order magnetic fluctuations (quartic fluctuations)~\cite{Zelle2024, Daviet2024, Hermansen2025Nov} together with the noise term encoded by the Keldysh component.

\subsection{Smaller interaction}
\label{subsec:Interaction}
\begin{figure}[t!]
\begin{center} 
\includegraphics[width=0.8\textwidth, angle=-0]{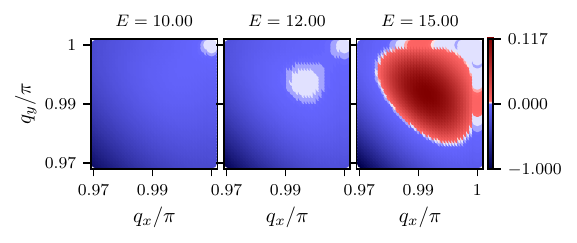}
\caption{
Momentum distribution of $\text{Im}\left[\left[\chi^{\bot, R}_{0,\bm{q},00}(\omega_q)\right]^{-1}\right]$ [a.u] for several drive amplitudes $E$, which are indicated above the each panel. All the other parameters are the same as in FIG. 4 of the main text except $I=4$.
}
\label{fig:polediagram_I4}
\end{center}
\end{figure}

In \Cref{fig:polediagram_I4}, we show a two dimensional false-color plot of the momentum dependence of $\text{Im}\left[\left[\chi^{\bot, R}_{0,\bm{q},00}(\omega_q)\right]^{-1}\right]$ for different drive strengths for linearly polarized radiation with polarization vector along the zone diagonal. Unlike FIG. 4 of the main text ($I=5$), here we use a smaller interaction strength $I=4$ to demonstrate that larger drive strength is required to induce the sign change of $\text{Im}\left[\left[\chi^{\bot, R}_{0,\bm{q},00}(\omega_q)\right]^{-1}\right]$.   

\bibliography{SI_Arxiv}%